\documentclass[11pt,letterpaper,twocolumn]{aastex63}
\usepackage{verbatim}
\usepackage{amsmath}

\usepackage{epsfig}
\usepackage{amsmath}
\usepackage{natbib}
\usepackage{mathtools}
\usepackage{longtable}

\begin{document}
 
\title{Linear Polarization Variations and Circular Polarization are Common Among Airless Bodies}
\author{Sloane J. Wiktorowicz}
\affiliation{Remote Sensing Department, The Aerospace Corporation, 2310 E. El Segundo Blvd., El Segundo, CA 90245}
\email{s.wiktorowicz@aero.org}

\author{Amanda J. Bayless}
\affiliation{Remote Sensing Department, The Aerospace Corporation, 2310 E. El Segundo Blvd., El Segundo, CA 90245}

\author{Larissa A. Nofi}
\affiliation{Remote Sensing Department, The Aerospace Corporation, 2310 E. El Segundo Blvd., El Segundo, CA 90245}

\date{\today}

\begin{abstract}

Using the POLISH2 polarimeter at the Lick Observatory Shane 3-m and Nickel 1-m telescopes, we discover rotation phase-locked variations in linear polarization to be common among asteroids and a NEO in a clear, 383 to 720 nm bandpass. Essentially all bodies in our eight-year study harbor statistically significant, repeatable linear polarization variations at the $0.01\% = 100$ ppm level or above (1 Ceres, 2 Pallas, 3 Juno, 4 Vesta, 6 Hebe, 7 Iris, 12 Victoria, 15 Eunomia, 16 Psyche, 132 Aethra, 216 Kleopatra, and 65803 Didymos). Since polarimetry is a differential technique, such variations cannot be due to shape changes and must be caused by heterogeneity in surface albedo and/or composition. While (4) Vesta has long been known to exhibit large, repeatable polarization variations across its surface, we discover the variations on (6) Hebe, (12) Victoria, and (65803) Didymos to be 1.5 to 3.5 times as large. We hypothesize that the \textit{DART} impact with Dimorphos blanketed Didymos with depolarizing ejecta, which suggests pristine variations across Didymos to have been even larger. As the only NEO in this study with data quality sufficient to investigate polarization variations, Didymos' huge variations suggest they may be common among NEOs. We also discover optical circular polarization to be enhanced for low-albedo, M type asteroids, which is correlated with their large radar albedos. Thus, we present optical circular polarimetry as an alternative method for the identification of metalliferous bodies. \\

\end{abstract}

\section{Introduction}\label{intro}
 
 Sunlight scattered by an airless surface imparts both linear and circular polarization that encodes a wealth of information about the surface. Briefly, polarization observables are given by Stokes $Q$, $U$ (for linear polarization), and $V$ (for circular polarization). These observables are normalized by Stokes $I$ (total intensity) to construct fractional polarization $q = Q/I$, $u = U/I$, and $v = V/I$, which are independent of object brightness or intensity variations caused by shape changes or atmospheric transparency. Linear polarization is given by $p = \sqrt{q^2 + u^2}$, and linear polarization orientation is given by the four-quadrant inverse tangent $\Theta = 1/2 \, \rm{atan}(u,q)$, which is typically denoted $\texttt{atan2}$ in computer languages. Here, $\Theta = 0^\circ$ indicates linear polarization oriented parallel to Celestial North-South.
 
\begin{figure*}
\centering
\includegraphics[width=0.65\textwidth]{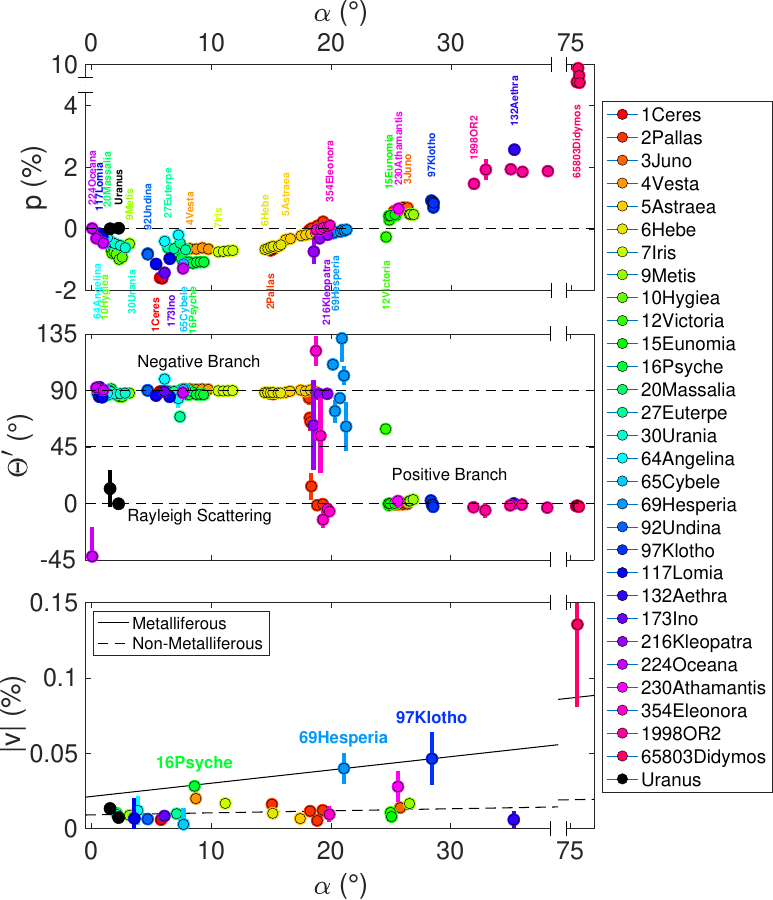}
\caption{Lick 3-m POLISH2 polarization phase curves of Main Belt Asteroids (MBAs), NEOs (1998 OR2 and 65803 Didymos), and Uranus (control target). Lick 1-m POLISH2 observations of (1) Ceres are also shown. Note the axis breaks to accommodate high phase angle observations of (65803) Didymos. \textit{Top:} Fractional linear polarization $p = P/I$ is plotted against phase angle $\alpha$, and linear polarization vanishes at the inversion angle $\alpha_0 \sim 20^\circ$. This angle is unique to each airless body and is tied to its mean surface index of refraction \citep{Muinonen2002a, Masiero2009, GilHutton2017}. Fractional polarization is defined to be negative for $\alpha < \alpha_0$ (``negative branch'') and positive for $\alpha > \alpha_0$ (``positive branch''). \textit{Middle:} Polarization orientation $\Theta'$ with respect to the plane perpendicular to the Sun-body-observer scattering plane. Here, the inversion angle $\alpha_0 \sim 20^\circ$ marks the discontinuity where polarization rapidly rotates by $\Delta \Theta' = \pm 90^\circ$ from parallel to the scattering plane (for $\alpha < \alpha_0$) to perpendicular to the scattering plane (Rayleigh-like, for $\alpha > \alpha_0$). The control target Uranus displays Rayleigh-like polarization orientation, perpendicular to the scattering plane ($\Theta' = 0.39^\circ \pm 0.72^\circ$), even down to $\alpha = 1.6^\circ$ due to its gaseous atmosphere. Thus, airless and gaseous bodies may be immediately distinguished from each other at low $\alpha$ simply from their polarization orientation $\Theta'$. \textit{Bottom}: Absolute value of fractional circular polarization $|v| = |V|/I$ vs. $\alpha$, where the metalliferous (16) Psyche, (69) Hesperia, and (97) Klotho harbor elevated circular polarization (see section \ref{sec_circ}). Linear fits to $|v|$ vs. $\alpha$ are shown for both the metalliferous sample (solid line) and the non-metalliferous sample (dashed line). The surface of the non-metalliferous Didymos may be contaminated by material excavated from Dimorphos during the \textit{DART} impact or may simply be a consequence of high phase angle observation.}
\label{phase}
\end{figure*}

 While not present in airless bodies, single scattering by gas molecules with sizes in the Rayleigh limit not only imparts a characteristic $\lambda^{-4}$ spectrum in scattered light, but it also imparts a polarization orientation that is perpendicular to the Sun-body-observer ``scattering plane'' ($\Theta' = 0^\circ$). It is straightforward to rotate polarization orientation $\Theta$ from the Celestial Frame to a frame perpendicular to the scattering plane. However, it has been known for nearly a century \citep{Lyot1929} that polarized backscattering of sunlight by airless bodies tends to impart polarization parallel to the scattering plane ($\Theta' = \pm 90^\circ$). For Sun-body-observer phase angles $\alpha < 20^\circ$ or thereabouts, both constructive and destructive interference are present for scattering geometries that generate polarization perpendicular to the Sun-body-observer scattering plane. However, only constructive interference is present for scattering geometries that generate polarization parallel to the scattering plane \citep{Muinonen2015}. Such ``coherent backscattering'' \citep{Shkuratov1988, Shkuratov1989, Muinonen1989, Muinonen1990, Muinonen2015} is a leading theory to describe why disk-integrated linear polarization of airless bodies lies on the ``negative branch.'' Here, polarization is oriented parallel to the scattering plane ($\Theta' = \pm 90^\circ$) and fractional linear polarization is defined to be $p < 0$. As phase angle increases to $\alpha \sim 20^\circ$, the ``inversion angle'' $\alpha_0$, linear polarization passes through zero and rotates from parallel to perpendicular to the scattering plane ($\Theta' = 0^\circ$, the ``positive branch''). Fractional linear polarization is then defined to be $p > 0$ (Figure \ref{phase}, top panel).

 The linear polarimetric behavior of all airless bodies may be described by three empirical parameters \citep{Masiero2012}: $\alpha_0$ (the ``inversion angle,'' which is the phase angle at which linear polarization vanishes), $p_{\rm min}$ (the largest absolute value of negative branch polarization, typically 1\% to 2\%), and $h$ (the linear slope of the polarization versus $\alpha$ trend at the inversion angle $\alpha_0$). The effect of these parameters on the polarization phase curve of airless bodies is shown in Figure \ref{phase}. At the inversion angle $\alpha_0 \sim 20^\circ$, polarization orientation $\Theta'$ rapidly rotates through $90^\circ$, causing $p$ to change sign. We have observed this rapid rotation to occur from one night to the next for (2) Pallas, (12) Victoria, (216) Kleopatra, and (354) Eleonora (Figure \ref{phase}, middle panel). The slope $h$ of the phase curve at the inversion angle is also diagnostic (e.g., 2 Pallas' steep slope compared to the shallower slopes of 5 Astraea and 354 Eleonora). Finally, the maximum extent of negative polarization $p_{\rm min}$ is quite large for (1) Ceres but small for (27) Euterpe (Figure \ref{phase}, top panel).
 
 Laboratory measurements of lunar fines, coarse rocks, and observations of asteroids occupy distinct phase space regions \citep{Dollfus1989} when plotting $p_{\rm min}$ versus $\alpha_0$. Indeed, asteroid polarization lies between the two end members of lunar fines and coarse rocks, which suggests that asteroid surfaces are dominated by relatively large regolith particles. In contrast, even century-old polarimetry of the Moon \citep{Lyot1929} suggested lunar regolith to be fine-grained, which was confirmed by samples from the Apollo Program. Therefore, polarimetric observations have significant potential for deeper understanding of airless body surfaces, as ground-based observations may be calibrated by \textit{in situ} measurements from spacecraft such as \textit{DART}, \textit{Dawn}, \textit{OSIRIS-REx}, the \textit{Psyche Mission}, and \textit{Rosetta}.
 
 Polarimetric observations are highly complementary to photometric ones. Since photometry is an absolute, non-differential quantity, changes in cross-sectional area (so-called ``shape changes'') are to first order degenerate with disk-integrated surface albedo variations as the airless body rotates. However, since polarimetry is inherently differential, it is insensitive to shape changes: as total flux changes with cross-sectional area, fractional polarization does not. Interestingly, the linear polarization and geometric albedo of rocky Solar System bodies has long been known to be anticorrelated via the so-called ``Umov effect'' \citep{Umov1905, Shkuratov1992, Zubko2011}: dark surfaces, dominated by single scattering, are strongly polarized, while multiple scattering in bright surfaces randomizes the electric field orientation and reduces linear polarization. Thus, rotational variability in linear polarization is a hallmark of surface heterogeneity on an airless body, and due to the strong photometric signature of shape changes, it is difficult to identify with spatially unresolved photometry alone. \\
 
\section{Observations}\label{}

\subsection{The POLISH2 Polarimeter} \label{instdesc}

POLISH2 (POlarimeter at Lick for Inclination Studies of Hot jupiters 2) is an aperture-integrated polarimeter that uses two photoelastic modulators (PEMs) instead of rotating waveplates to achieve part-per-million (ppm) sensitivity and accuracy on an incident beam \citep{WiktorowiczNofi2015, Wiktorowicz2023}. The POLISH2 PEMs are fused silica bars piezoelectrically resonated at their fundamental modes. After passing the beam through a Wollaston prism, the time-varying stress birefringence from the PEMs AC-couples incident polarization into intensity modulations at 10 to 100 kHz that are detected by photomultiplier tube modules. POLISH2's high sensitivity and accuracy are driven by the highly linear response of PEMs to incident polarization as well as their high frequency modulation. Thus, POLISH2 may be thought of as a polarimeter with a waveplate rotating at $\sim 50$ kHz, but its suppression of systematic effects is far superior to that of waveplate polarimeters \citep{Wiktorowicz2023}. Indeed, POLISH2 conclusively detected scattered light from the hot Jupiter exoplanet HD 189733b with ppm accuracy independently from both the Lick Observatory 3-m and the Gemini North 8-m \citep{Wiktorowicz2025}.

Dominant systematic effects include polarized background sky, especially in moonlit conditions, and telescope polarization. Subtraction of sky polarization is accomplished by the automated POLISH2 observing cadence, where 30 sec of on-target data are acquired followed by 30 sec on a sky field 30 arcsec due North and finally 30 sec on-target. Observations of a particular target include many such observing ``triplets.'' Telescope polarization contamination, typically $\sim 0.007\% = 70$ ppm at the Lick 3-m and measured to be stable to within $\sim 10$ ppm over a ten-year period \citep{Wiktorowicz2023}, is corrected by observation of a large amount of weakly polarized stars. Such stars are identified by Lick 1-m POLISH2 surveys of bright, nearby stars expected to harbor minimal interstellar polarization. Strongly polarized calibrator stars enable repeatable rotational zero point accuracy of $\sigma_\Theta = -1.08^\circ \pm 0.34^\circ$ from run to run \citep{WiktorowiczNofi2015}. Further detail regarding the POLISH2 instrument, its operation, and its calibration may be found in \cite{Wiktorowicz2023}.

\subsection{Linear Polarization Variations} \label{linpolvar}

Using our POLISH2 polarimeter at the Lick 3-m telescope \citep{WiktorowiczNofi2015, Wiktorowicz2023}, with observations of (1) Ceres obtained at the Lick 1-m telescope, we have obtained a significant amount of data on airless bodies from UT 20 Apr 2014 to 25 Oct 2022 to search for rotation phase-locked linear polarization variations. The brightest observable bodies are chosen for each observing run to maximize measurement SNR, and measured linear polarization is phased to rotation periods obtained from the JPL Small-Body Database. The zero point of this rotational phase is arbitrary. Nightly mean data are plotted in Figure \ref{phase} and listed in Table \ref{alldata}, where quantities in parenthesis indicate $1 \sigma$ uncertainties in the last two digits of each parameter. Parenthetical quantities in phase angle $\alpha$ denote half of the width of each phase angle bin. Stokes $q'$, $u'$, and $\Theta'$ result from rotating unprimed Stokes $q$, $u$, and $\Theta$ from the celestial frame to perpendicular to the scattering plane, where scattering plane orientation is obtained from the JPL Horizons database. Note fractional linear polarization Stokes $p$ and circular polarization Stokes $v$ are invariant under rotation. Most Stokes $v$ values are binned over each weeklong observing run to increase SNR.

\newpage
\startlongtable
\begin{deluxetable*}{cccccccccc}
\tabletypesize{\tiny}
\tablecaption{POLISH2 Nightly Linear and Circular Polarimetry of Solar System Objects}
\tablewidth{0pt}
\tablehead{
\colhead{Object} & \colhead{UT Date} & \colhead{$\alpha$ ($^\circ$)}	& \colhead{$q'$ (\%)} & \colhead{$u'$ (\%)}	& \colhead{$v$ (\%)}		& \colhead{$p$ (\%)}		& \colhead{$\Theta'$ ($^\circ$)}}
\startdata
(1) Ceres        & 2014 Apr 20 	 & 5.7  	 & $-$1.583(34)     	 & 0.0521(20)  	 & $-$  	 		& $-$1.584(34)   	 & 89.058(41) \\
$\cdots$        & 2014 Apr 21 	 & 5.9  	 & $-$1.600(32)     	 & 0.0552(17)  	 & $-$		   	 & $-$1.601(32)   	 & 89.012(37) \\
$\cdots$        & 2014 Apr 20-21			& 5.811(99)	& $-$ 	& $-$ 			& 0.0059(30) 		& $-$ 			& $-$    \\
\hline
(2) Pallas       & 2015 Feb 10 	 & 18.2 	 & $-$0.0376(76)    	 & 0.0081(71)  	 & $-$  			 & $-$0.0378(76)  	 & 83.9(5.4)  \\
$\cdots$       & 2015 Feb 11 	 & 18.2 	 & $-$0.066(11)     	 & 0.063(10)   	 	& $-$   			 & $-$0.090(11)   	 & 68.1(3.4) \\
$\cdots$       & 2015 Feb 12 	 & 18.3 	 & $-$0.0238(94)    	 & 0.0286(87)  	 & $-$     	 		& $-$0.0360(90)  	 & 64.9(7.2) \\
$\cdots$       & 2015 Feb 13 	 & 18.4 	 & 0.0189(80) 	 & 0.0098(73)  	 & $-$		     	 & 0.0200(79)     	 & 14(11)      \\
$\cdots$       & 2015 Feb 10-13	& 18.27(10)    	 & $-$		& $-$			& $-$0.0116(41)  	& $-$			& $-$ \\
$\cdots$       & 2017 Sep 09 	 & 19.3 	 & 0.2360(15) 	 & $-$0.0045(12)     	 & 0.0122(13)     	 & 0.2361(15)     	 & $-$0.55(14)    \\
$\cdots$       & 2019 May 23 	 & 18.9 	 & 0.1053(28) 	 & $-$0.0053(36)     	 & $-$0.0054(32)  	 & 0.1053(29)     	 & $-$1.44(98)    \\
$\cdots$       & 2020 Feb 23 	 & 14.9 	 & $-$0.6919(49)    	 & 0.0636(46)  	 & $-$		  	 & $-$0.6948(49)  	 & 87.38(19)  \\
$\cdots$       & 2020 Feb 24 	 & 15.0 	 & $-$0.6767(52)    	 & 0.0650(48)  	 & $-$  			 & $-$0.6798(52)  	 & 87.26(20)  \\
$\cdots$       & 2020 Feb 26 	 & 15.2 	 & $-$0.6316(57)    	 & 0.0659(51)  	 & $-$		  	 & $-$0.6350(57)  	 & 87.02(23)  \\
$\cdots$       & 2020 Feb 23-26	& 15.08(14)    	 & $-$		& $-$			& $-$0.0162(30)	& $-$			& $-$  \\
\hline
(3) Juno         & 2018 Sep 18 	 & 26.4 	 & 0.6876(40) 	 & $-$0.0123(32)     	 & $-$		     	 & 0.6877(40)     	 & $-$0.51(13)    \\
$\cdots$         & 2018 Sep 19 	 & 26.2 	 & 0.6710(46) 	 & $-$0.0020(28)     	 & $-$		  	 & 0.6710(46)     	 & $-$0.09(12)    \\
$\cdots$         & 2018 Sep 20 	 & 26.0 	 & 0.6971(53) 	 & $-$0.0181(43)     	 & $-$		  	 & 0.6974(53)     	 & $-$0.74(18)    \\
$\cdots$         & 2018 Sep 21 	 & 25.8 	 & 0.6480(28) 	 & $-$0.0041(23)     	 & $-$		  	 & 0.6480(28)     	 & $-$0.18(10)  \\
$\cdots$         & 2018 Sep 22 	 & 25.6 	 & 0.6208(51) 	 & $-$0.0176(30)     	 & $-$		  	 & 0.6210(51)     	 & $-$0.81(14)    \\
$\cdots$         & 2018 Sep 23 	 & 25.4 	 & 0.6034(30) 	 & $-$0.0147(18)     	 & $-$		    	 & 0.6036(30)     	 & $-$0.697(84)   \\
$\cdots$         & 2018 Sep 24 	 & 25.2 	 & 0.5786(23) 	 & $-$0.0145(18)     	 & $-$		  	 & 0.5787(23)     	 & $-$0.718(88)   \\
$\cdots$         & 2018 Sep 18-24  & 25.80(58) 	 & $-$	 	 & $-$		     	 & 0.01391(84)  	 & $-$     	 		& $-$   \\
\hline
(4) Vesta        & 2018 Jun 02 	 & 9.7  	 & $-$0.65201(56)   	 & $-$0.01768(47)    	 & $-$		 	 & $-$0.65225(56) 	 & 90.777(21)    \\
$\cdots$        & 2018 Jun 03 	 & 9.2  	 & $-$0.61440(43)   	 & $-$0.02250(38)    	 & $-$		 	 & $-$0.61481(43) 	 & 91.049(18)    \\
$\cdots$        & 2018 Jun 04 	 & 8.7  	 & $-$0.65074(35)   	 & $-$0.01758(30)    	 & $-$		 	 & $-$0.65097(35) 	 & 90.774(13)    \\
$\cdots$        & 2018 Jun 05 	 & 8.2  	 & $-$0.64146(45)   	 & $-$0.02453(39)    	 & $-$		    	 & $-$0.64193(45) 	 & 91.095(17)    \\
$\cdots$        & 2018 Jun 06 	 & 7.7  	 & $-$0.66824(63)   	 & $-$0.02624(55)    	 & $-$		    	 & $-$0.66875(63) 	 & 91.124(23)    \\
$\cdots$         & 2018 Jun 02-06  & 8.7(1.0) & $-$	 	 	& $-$		     	 & 0.01999(17)  	 & $-$     	 		& $-$   \\
\hline
(5) Astraea      & 2020 Feb 20 	 & 16.2 	 & $-$0.3521(80)    	 & 0.0213(67)  	 & $-$		  	 & $-$0.3527(80)  	 & 88.27(55)  \\
$\cdots$      & 2020 Feb 21 	 & 16.6 	 & $-$0.3158(44)    	 & 0.0075(37)  	 & $-$		  	 & $-$0.3159(44)  	 & 89.32(33)  \\
$\cdots$      & 2020 Feb 23 	 & 17.5 	 & $-$0.2209(82)    	 & $-$0.0012(68)     	 & $-$		  	 & $-$0.2209(82)  	 & 90.16(89)     \\
$\cdots$      & 2020 Feb 24 	 & 17.9 	 & $-$0.1918(85)    	 & 0.0038(71)  	 & $-$		  	 & $-$0.1918(85)  	 & 89.4(1.1)  \\
$\cdots$      & 2020 Feb 25 	 & 18.3 	 & $-$0.1626(22)    	 & $-$0.0015(18)     	 & $-$		  	 & $-$0.1626(22)  	 & 90.26(32)     \\
$\cdots$      & 2020 Feb 26 	 & 18.7 	 & $-$0.1339(79)    	 & 0.0041(65)  	 & $-$		  	 & $-$0.1339(79)  	 & 89.1(1.4)  \\
$\cdots$       & 2020 Feb 20-26  & 17.4(1.2) & $-$	 	 	& $-$		     	 & $-$0.0067(20)  	 & $-$     	 		& $-$   \\
\hline
(6) Hebe         & 2020 Feb 20 	 & 15.8 	 & $-$0.5126(27)    	 & 0.0382(23)  	 & $-$		  	 & $-$0.5141(27)  	 & 87.87(13)  \\
$\cdots$         & 2020 Feb 21 	 & 15.6 	 & $-$0.5653(37)    	 & 0.0407(31)  	 & $-$		  	 & $-$0.5668(37)  	 & 87.94(16)  \\
$\cdots$         & 2020 Feb 23 	 & 15.2 	 & $-$0.5625(29)    	 & 0.0348(25)  	 & $-$		 	 & $-$0.5636(29)  	 & 88.23(12)  \\
$\cdots$         & 2020 Feb 24 	 & 14.9 	 & $-$0.5891(41)    	 & 0.0408(34)  	 & $-$		  	 & $-$0.5905(41)  	 & 88.02(17)  \\
$\cdots$         & 2020 Feb 25 	 & 14.7 	 & $-$0.6296(23)    	 & 0.0417(19)  	 & $-$		  	 & $-$0.6309(23)  	 & 88.103(87) \\
$\cdots$         & 2020 Feb 26 	 & 14.5 	 & $-$0.6632(30)    	 & 0.0425(25)  	 & $-$		  	 & $-$0.6645(30)  	 & 88.17(11)  \\
$\cdots$         & 2020 Feb 20-26  & 15.14(67) & $-$	 	 	& $-$		     	 & $-$0.0102(13)  	 & $-$     	 		& $-$   \\
\hline
(7) Iris         & 2015 Feb 10 	 & 11.8 	 & $-$0.6993(43)    	 & 0.0041(52)  	 & $-$		  	 & $-$0.6993(43)  	 & 89.83(21)  \\
$\cdots$         & 2015 Feb 11 	 & 11.4 	 & $-$0.7090(29)    	 & 0.0091(36)  	 & $-$		     	 & $-$0.7090(29)  	 & 89.63(14)  \\
$\cdots$         & 2015 Feb 12 	 & 11.0 	 & $-$0.7306(26)    	 & 0.0121(32)  	 & $-$		     	 & $-$0.7307(26)  	 & 89.53(13)  \\
$\cdots$         & 2015 Feb 13 	 & 10.6 	 & $-$0.7430(23)    	 & 0.0090(28)  	 & $-$		     	 & $-$0.7430(23)  	 & 89.65(11)  \\
$\cdots$         & 2015 Feb 10-13  & 11.18(64) & $-$	 	 	& $-$		     	 & $-$0.0168(15)  	 & $-$     	 		& $-$   \\
\hline
(9) Metis        & 2019 Oct 28 	 & 3.2  	 & $-$0.48139(78)   	 & 0.03667(65) 	 & $-$0.00887(72)	& $-$0.48278(78) 	 & 87.822(39) \\
$\cdots$        & 2020 Feb 21 	 & 26.9 	 & 0.465(26)  	 & 0.049(24)   	 & $-$		      	 & 0.467(26)      	 & 3.0(1.5)    \\
$\cdots$        & 2020 Feb 24 	 & 26.6 	 & 0.4774(90) 	 & 0.0330(83)  	 & $-$		 	 & 0.4785(90)     	 & 1.98(50)    \\
$\cdots$        & 2020 Feb 25 	 & 26.5 	 & 0.4782(35) 	 & 0.0296(32)  	 & $-$		  	 & 0.4791(35)     	 & 1.77(19)    \\
$\cdots$         & 2020 Feb 21-25  & 26.568(50) & $-$	 	 	& $-$		     	 & $-$0.0166(31)  	 & $-$     	 		& $-$   \\
\hline
(10) Hygiea      & 2018 Sep 18 	 & 2.6  	 & $-$0.9019(43)    	 & 0.1591(46)  	 & $-$		    	 & $-$0.9158(43)  	 & 85.00(14)  \\
$\cdots$      & 2018 Sep 19 	 & 2.3  	 & $-$0.9682(54)    	 & 0.1674(53)  	 & $-$		    	 & $-$0.9825(54)  	 & 85.10(15)  \\
$\cdots$      & 2018 Sep 20 	 & 2.1  	 & $-$0.8116(47)    	 & 0.0940(42)  	 & $-$		    	 & $-$0.8170(47)  	 & 86.70(15)  \\
$\cdots$      & 2018 Sep 21 	 & 1.9  	 & $-$0.7483(44)    	 & 0.0574(35)  	 & $-$		    	 & $-$0.7505(44)  	 & 87.81(13)  \\
$\cdots$      & 2018 Sep 22 	 & 1.7  	 & $-$0.7877(68)    	 & 0.0051(40)  	 & $-$		    	 & $-$0.7878(68)  	 & 89.82(15)  \\
$\cdots$      & 2018 Sep 23 	 & 1.6  	 & $-$0.6807(67)    	 & $-$0.0293(51)     	 & $-$		    	 & $-$0.6813(67)  	 & 91.23(21)     \\
$\cdots$         & 2018 Sep 18-23  & 2.10(46) & $-$	 	 	& $-$		     	 & 0.0104(15)  		 & $-$     	 		& $-$   \\
\hline
(12) Victoria    & 2021 Sep 14 	 & 24.6 	 & $-$0.1244(20)    	 & 0.2286(23)  	 & $-$		  	 & $-$0.2603(22)  	 & 59.28(23) \\
$\cdots$    & 2021 Sep 15 	 & 24.8 	 & 0.2897(21) 	 & $-$0.0157(16)     	 & $-$		  	 & 0.2901(21)     	 & $-$1.56(16)    \\
$\cdots$    & 2021 Sep 16 	 & 25.1 	 & 0.4464(24) 	 & $-$0.0087(19)     	 & $-$		  	 & 0.4465(24)     	 & $-$0.56(12)    \\
$\cdots$    & 2021 Sep 17 	 & 25.4 	 & 0.4389(23) 	 & $-$0.0072(18)     	 & $-$		  	 & 0.4390(23)     	 & $-$0.47(12)    \\
$\cdots$         & 2021 Sep 14-17  & 24.99(42) & $-$	 	 	& $-$		     	 & $-$0.01076(98)  	& $-$     	 		& $-$   \\
\hline
(15) Eunomia     & 2019 Oct 23 & 24.8 	 & 0.4187(17) 	 & $-$0.0041(15)     & $-$		       	 & 0.4188(17)     	 & $-$0.28(10)    \\
$\cdots$     & 2019 Oct 24 	 & 24.9 	 & 0.4310(20) 	 & $-$0.0036(15)     	 & $-$		      	 & 0.4310(20)     	 & $-$0.237(98)   \\
$\cdots$     & 2019 Oct 28 	 & 25.3 	 & 0.4684(15) 	 & 0.0002(14)  	 & $-$		       	 & 0.4684(15)     	 & 0.011(84)   \\
$\cdots$         & 2019 Oct 23-28  & 25.07(26) & $-$	 	 	& $-$		     	 & $-$0.00799(75)  	& $-$     	 		& $-$   \\
\hline
(16) Psyche      & 2018 May 31 	 & 7.5  	 & $-$0.95(12)      	 & $-$0.02(13)       	 	& $-$		       	 & $-$0.95(12)    	 & 90.5(4.0)     \\
$\cdots$      & 2018 Jun 01 	 & 7.8  	 & $-$1.0956(37)    	 & 0.1131(39)  	 & $-$		      	 & $-$1.1014(37)  	 & 87.05(10)  \\
$\cdots$      & 2018 Jun 02 	 & 8.1  	 & $-$1.0661(25)    	 & 0.1282(26)  	 & $-$		  	 & $-$1.0737(25)  	 & 86.572(70) \\
$\cdots$      & 2018 Jun 03 	 & 8.5  	 & $-$1.1121(26)    	 & 0.0513(27)  	 & $-$		     	 & $-$1.1133(26)  	 & 88.680(69) \\
$\cdots$      & 2018 Jun 04 	 & 8.8  	 & $-$1.0763(28)    	 & 0.1219(28)  	 & $-$		   	 & $-$1.0832(28)  	 & 86.770(75) \\
$\cdots$      & 2018 Jun 05 	 & 9.1  	 & $-$1.0651(25)    	 & 0.1222(25)  	 & $-$		   	 & $-$1.0721(25)  	 & 86.727(68) \\
$\cdots$      & 2018 Jun 06 	 & 9.4  	 & $-$1.0662(30)    	 & 0.1258(30)  	 & $-$		   	 & $-$1.0736(30)  	 & 86.635(81) \\
$\cdots$         & 2018 May 31-Jun 06  & 8.60(79) & $-$	 	& $-$		     	 & 0.0282(11)  	& $-$     	 		& $-$   \\
\hline
(20) Massalia    & 2019 May 23 	 & 1.3  	 & $-$0.387(14)     	 & 0.053(14)   	 & $-$		   	 & $-$0.390(14)   	 & 86.1(1.1)  \\
\hline
(27) Euterpe     & 2018 Sep 18 	 & 6.4  	 & $-$0.6111(21)    	 & 0.0129(28)  	 & $-$		     	 & $-$0.6112(21)  	 & 89.39(13)  \\
$\cdots$     & 2018 Sep 19 	 & 6.9  	 & $-$0.6397(22)    	 & 0.0182(29)  	 & $-$		     	 & $-$0.6400(22)  	 & 89.19(13)  \\
$\cdots$     & 2018 Sep 20 	 & 7.4  	 & $-$0.3441(66)    	 & 0.309(11)   	 & $-$		  	 & $-$0.4625(86)  	 & 69.03(56) \\
$\cdots$     & 2018 Sep 21 	 & 7.8  	 & $-$0.6755(24)    	 & $-$0.0238(26)     	 & $-$		  	 & $-$0.6760(24)  	 & 91.01(11)     \\
$\cdots$         & 2018 Sep 18-21  & 7.11(76) & $-$	 	 	& $-$		     	 & 0.00987(92)  	& $-$     	 		& $-$   \\
\hline
(30) Urania      & 2018 Sep 18 	 & 1.5  	 & $-$0.4558(40)    	 & $-$0.0066(33)     	 & $-$		     	 & $-$0.4559(40)  	 & 90.42(21)     \\
$\cdots$      & 2018 Sep 19 	 & 1.4  	 & $-$0.3829(39)    	 & 0.0008(51)  	 & $-$		  	 & $-$0.3829(39)  	 & 89.94(38)  \\
$\cdots$      & 2018 Sep 20 	 & 1.6  	 & $-$0.4154(33)    	 & 0.0140(41)  	 & $-$		  	 & $-$0.4156(33)  	 & 89.03(28)  \\
$\cdots$      & 2018 Sep 21 	 & 1.9  	 & $-$0.4734(41)    	 & 0.0494(42)  	 & $-$		     	 & $-$0.4759(41)  	 & 87.02(26)  \\
$\cdots$      & 2018 Sep 22 	 & 2.4  	 & $-$0.5345(56)    	 & 0.0506(41)  	 & $-$		     	 & $-$0.5369(55)  	 & 87.30(22)  \\
$\cdots$      & 2018 Sep 23 	 & 2.8  	 & $-$0.6082(62)    	 & 0.0480(39)  	 & $-$		     	 & $-$0.6100(62)  	 & 87.75(19)  \\
$\cdots$         & 2018 Sep 18-23  & 2.12(69) & $-$	 	 	& $-$		     	 & 0.0095(13)  	& $-$     	 		& $-$   \\
\hline
(64) Angelina    & 2021 Sep 14 	 & 0.5  	 & $-$0.179(11)     	 & 0.014(12)   	 &  $-$		      	 & $-$0.180(11)   	 & 87.7(1.8)  \\
$\cdots$    & 2021 Sep 15 	 & 0.6  	 & $-$0.190(11)     	 & $-$0.005(11)      	 & $-$		   	 & $-$0.190(11)   	 & 90.7(1.7)     \\
$\cdots$    & 2021 Sep 16 	 & 1.0  	 & $-$0.236(13)     	 & 0.039(10)   	 & $-$		   	 & $-$0.239(13)   	 & 85.4(1.2)  \\
$\cdots$    & 2021 Sep 17 	 & 1.3  	 & $-$0.281(14)     	 & 0.022(11)   	 &  $-$		      	 & $-$0.282(14)   	 & 87.7(1.1)  \\
$\cdots$    & 2021 Sep 29 	 & 6.1  	 & $-$0.384(59)     	 & $-$0.125(55)      	 & $-$		   	 & $-$0.400(59)   	 & 99.0(4.0)     \\
$\cdots$    & 2021 Oct 02 	 	& 7.3  	 & $-$0.196(51)     	 & 0.046(49)   	 &  $-$		      	 & $-$0.196(51)   	 & 83.4(7.1)  \\
$\cdots$         & 2021 Sep 14-Oct 02  & 3.9(3.4) & $-$	 & $-$		     	 & $-$0.0121(95)  	& $-$     	 		& $-$   \\
\hline
(65) Cybele      & 2022 Oct 24 	 & 7.7  	 & $-$1.131(10)     	 & $-$0.033(11)      	 & 0.003(11)      	 & $-$1.131(10)   	 & 90.85(29)     \\
\hline
(69) Hesperia    & 2020 Feb 20 	 & 20.2 	 & $-$0.102(20)     	 & $-$0.090(16)      	 & $-$		   	 & $-$0.135(18)   	 & 110.6(3.8)    \\
$\cdots$    & 2020 Feb 21 	 & 20.3 	 & $-$0.090(37)     	 & 0.059(30)   	 & $-$		   	 & $-$0.103(35)   	 & 73.5(9.1) \\
$\cdots$    & 2020 Feb 23 	 & 20.8 	 & $-$0.088(19)     	 & 0.019(15)   	 & $-$		   	 & $-$0.089(19)   	 & 83.8(5.0)  \\
$\cdots$    & 2020 Feb 24 	 & 20.9 	 & $-$0.004(20)     	 & $-$0.032(16)      	 & $-$		   	 & $-$0.032(16)   	 & 131(19)       \\
$\cdots$    & 2020 Feb 25 	 & 21.1 	 & $-$0.054(17)     	 & $-$0.023(14)      	 & $-$0.040(10)		   	 & $-$0.057(17)   	 & 101.7(7.5)    \\
$\cdots$    & 2020 Feb 26 	 & 21.3 	 & $-$0.015(17)     	 & 0.023(14)   	 & $-$		      	 & $-$0.024(14)   	 & 61(20)    \\
\hline
(92) Undina      & 2022 Oct 24 	 & 4.7  	 & $-$0.7969(74)    	 & $-$0.0022(70)     	 & $-$		  	 & $-$0.7969(74)  	 & 90.08(25)     \\
$\cdots$      & 2022 Oct 25 	 & 4.7  	 & $-$0.8225(57)    	 & 0.0036(50)  	 & $-$		     	 & $-$0.8225(57)  	 & 89.87(18)  \\
$\cdots$         & 2022 Oct 24-25  & 4.6907(34) & $-$	 	 	& $-$		     	 & $-$0.0064(42)  	& $-$     	 		& $-$   \\
\hline
(97) Klotho      & 2020 Feb 21 	 & 28.4 	 & 0.918(53)  	 & 0.074(44)   	 & $-$		      	 & 0.920(52)      	 & 2.3(1.4)    \\
$\cdots$      & 2020 Feb 23 	 & 28.5 	 & 0.876(24)  	 & $-$0.036(20)      	 & $-$		      	 & 0.877(24)      	 & $-$1.17(66)    \\
$\cdots$      & 2020 Feb 24 	 & 28.5 	 & 0.874(75)  	 & $-$0.015(63)      	 & $-$		      	 & 0.874(75)      	 & $-$0.5(2.1)    \\
$\cdots$      & 2020 Feb 25 	 & 28.5 	 & 0.698(26)  	 & $-$0.025(21)      	 & $-$		      	 & 0.698(26)      	 & $-$1.01(88)    \\
$\cdots$      & 2020 Feb 26 	 & 28.5 	 & 0.833(23)  	 & $-$0.079(19)      	 & $-$		   	 & 0.837(23)      	 & $-$2.69(67)    \\
$\cdots$         & 2020 Feb 21-26  & 28.5 	& $-$	 	 	& $-$		     	 & 0.046(18)  	& $-$     	 		& $-$   \\
\hline
(117) Lomia      & 2021 Sep 14 	 & 0.9  	 & $-$0.247(14)     	 & 0.048(15)   	 & $-$		      	 & $-$0.251(14)   	 & 84.5(1.7)  \\
$\cdots$      & 2021 Sep 15 	 & 0.7  	 & $-$0.155(16)     	 & $-$0.013(12)      	 & $-$		   	 & $-$0.155(16)   	 & 92.5(2.3)     \\
$\cdots$      & 2021 Sep 16 	 & 0.6  	 & $-$0.171(13)     	 & 0.030(16)   	 & $-$		      	 & $-$0.172(13)   	 & 85.0(2.6)  \\
$\cdots$      & 2021 Sep 17 	 & 0.9  	 & $-$0.245(14)     	 & 0.038(15)   	 & $-$		   	 & $-$0.247(14)   	 & 85.6(1.8)  \\
$\cdots$      & 2021 Sep 29 	 & 5.4  	 & $-$1.13(10)      	 & 0.164(91)   	 & $-$		   	 & $-$1.14(10)    	 & 85.9(2.3)  \\
$\cdots$      & 2021 Oct 02 	 & 6.5  	 & $-$0.951(65)     	 & 0.170(57)   	 & $-$		      	 & $-$0.964(65)   	 & 84.9(1.7)  \\
$\cdots$         & 2021 Sep 14-Oct 02  & 3.6(2.9)	& $-$	& $-$		 & 0.007(14)  		& $-$     	 		& $-$   \\
\hline
(132) Aethra     & 2020 Feb 26 	 & 35.3 	 & 2.5730(61) 	 & $-$0.0075(51)     	 & $-$0.0060(56)  	 & 2.5730(61)     	 & $-$0.083(57)   \\
\hline
(173) Ino        & 2018 Sep 22 	 & 6.1  	 & $-$1.4191(39)    	 & 0.0489(22)  	 & 0.0084(15)  	 & $-$1.4200(39)  	 & 89.014(44) \\
\hline
(216) Kleopatra  & 2017 Sep 08 	 & 18.5 	 & $-$0.48(93)      	 & 0.70(35)    & $-$		       	 & $-$0.73(40)    	 & 62(36)    \\
$\cdots$  & 2017 Sep 10 	 & 19.0 	 & $-$0.3028(47)    	 & 0.0312(36)  	 & $-$		    	 & $-$0.3044(47)  	 & 87.06(34)  \\
$\cdots$  & 2017 Sep 13 	 & 19.7 	 & $-$0.1890(49)    	 & 0.0183(39)  	 & $-$		    	 & $-$0.1899(49)  	 & 87.23(59)  \\
\hline
(224) Oceana     & 2021 Sep 14 	 & 0.4  	 & $-$0.305(13)     	 $-$& 0.022(12)      & $-$		  	 & $-$0.306(13)   	 & 92.1(1.1)     \\
$\cdots$     & 2021 Sep 15 	 & 0.1  	 & 0.002(11)  & $-$0.017(13)      	 & $-$		      	 & 0.017(13)      	 & $-$42(24)       \\
$\cdots$     & 2021 Sep 17 	 & 1.0  	 & $-$0.453(10)     	 & $-$0.005(11)      	 & $-$		   	 & $-$0.453(10)   	 & 90.32(67)     \\
$\cdots$     & 2021 Oct 02 	 & 7.6  	 & $-$1.279(67)     	 & 0.085(66)   	 & $-$		   	 & $-$1.280(67)   	 & 88.1(1.5)  \\
\hline
(230) Athamantis & 2018 Jun 02 	 & 25.6 	 & 0.656(11)  	 & 0.044(11)   	 & $-$0.028(11)   	 & 0.657(11)      	 & 1.92(49)    \\
\hline
(354) Eleonora   & 2020 Feb 20 	 & 19.9 	 & 0.112(13)  	 & $-$0.025(11)       & $-$0.0095(53)   	 & 0.115(13)      & $-$6.3(2.7)    \\
$\cdots$   & 2020 Feb 21 	 & 19.7 	 & 0.075(12)  	 & $-$0.0113(98)     	 & $-$		   	 & 0.075(12)      	 & $-$4.3(3.7)    \\
$\cdots$   & 2020 Feb 23 	 & 19.3 	 & 0.043(12)  	 & $-$0.021(10)      	 & $-$		   	 & 0.046(12)      	 & $-$12.9(6.5)    \\
$\cdots$   & 2020 Feb 24 	 & 19.1 	 & $-$0.006(17)     	 & 0.019(14)   	 	& $-$		   	 & $-$0.019(14)   	 & 54(30)    \\
$\cdots$   & 2020 Feb 26 	 & 18.8 	 & $-$0.013(11)     	 & $-$0.0251(93)     	 	& $-$		      	 & $-$0.0261(97)  	 & 121(12)       \\
\hline
1998 OR2 & 2020 Feb 20 	 & 31.9 	 & 1.457(56)  	 & $-$0.158(69)      	 & $-$		      	 & 1.464(56)      	 & $-$3.1(1.3)    \\
$\cdots$ & 2020 Feb 21 	 & 32.9 	 & 1.92(35)   	 & $-$0.38(39)       	 	& $-$		    	 & 1.92(35)       	 	& $-$5.5(5.8)    \\
$\cdots$ & 2020 Feb 23 	 & 35.0 	 & 1.936(58)  	 & $-$0.108(70)      	 & $-$		   	 & 1.938(58)      	 & $-$1.6(1.0)    \\
$\cdots$ & 2020 Feb 24 	 & 36.0 	 & 1.853(50)  	 & $-$0.066(59)      	 & $-$		      	 & 1.853(50)      	 & $-$1.02(91)    \\
$\cdots$ & 2020 Feb 26 	 & 38.1 	 & 1.864(56)  	 & $-$0.222(65)      	 & $-$		   	 & 1.876(56)      	 & $-$3.39(99)    \\
\hline
(65803) Didymos  & 2022 Oct 20 	 & 76.3 	 & 9.31(13)   & $-$0.71(11)       	 & $-$		    	 & 9.33(13)       	 & $-$2.17(34)    \\
$\cdots$  		& 2022 Oct 21 	 & 76.3 	 & 8.86(10)   	 & $-$0.826(87)      & $-$		   	 & 8.90(10)       	 & $-$2.66(28)    \\
$\cdots$  		& 2022 Oct 24 	 & 76.1 	 & 9.76(13)   	 & $-$0.62(11)       	 & $-$		   	 & 9.78(13)       	 & $-$1.81(31)    \\
$\cdots$  		& 2022 Oct 25 	 & 75.9 	 & 8.91(13)   	 & $-$0.67(11)       	 & $-$	       	 	& 8.94(13)       	 & $-$2.14(36)    \\
$\cdots$         & 2022 Oct 20-25  & 76.13(22)	& $-$	& $-$		 	& $-$0.136(55)  	& $-$     	 		& $-$   \\
\hline
Uranus        & 2018 Sep 22 	 & 1.6  	 & 0.0062(34) 	 & 0.0027(30)  	 & 0.0134(16)  	 	& 0.0062(33)     	 & 12(15)      \\
$\cdots$        & 2021 Sep 16 	 & 2.3  	 & 0.0246(34) 	 & $-$0.0003(31)     	 & $-$0.0074(31)     	 & 0.0246(34)     	 & $-$0.3(3.6) 
\label{alldata}
\enddata
\end{deluxetable*}

\begin{figure}
\centering
\includegraphics[width=0.47\textwidth]{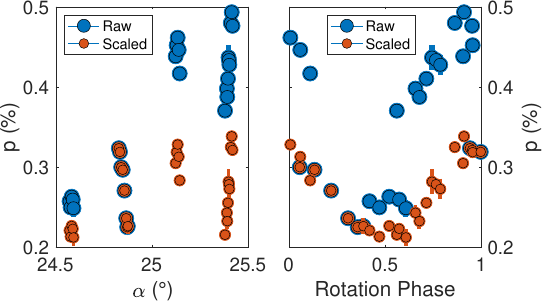}
\caption{\textit{Left}: Fractional linear polarization $p$ versus $\alpha$ for (12) Victoria obtained on UT 14 to 17 Sep 2021 with Lick 3-m POLISH2. The polarization phase curve varies both from nightly changes in $\alpha$ (blue points) and from intrinsic rotational variations across the asteroid's surface (red points). \textit{Right}: (12) Victoria observations phased to the 8.66 h rotation period of the asteroid. Observations made at overlapping rotational phases are used to subtract polarization variations due to phase angle changes and uncover variations intrinsic to the asteroid (red points).}
\label{phasecurve}
\end{figure}

\begin{figure*}
\centering
\includegraphics[width=0.75\textwidth]{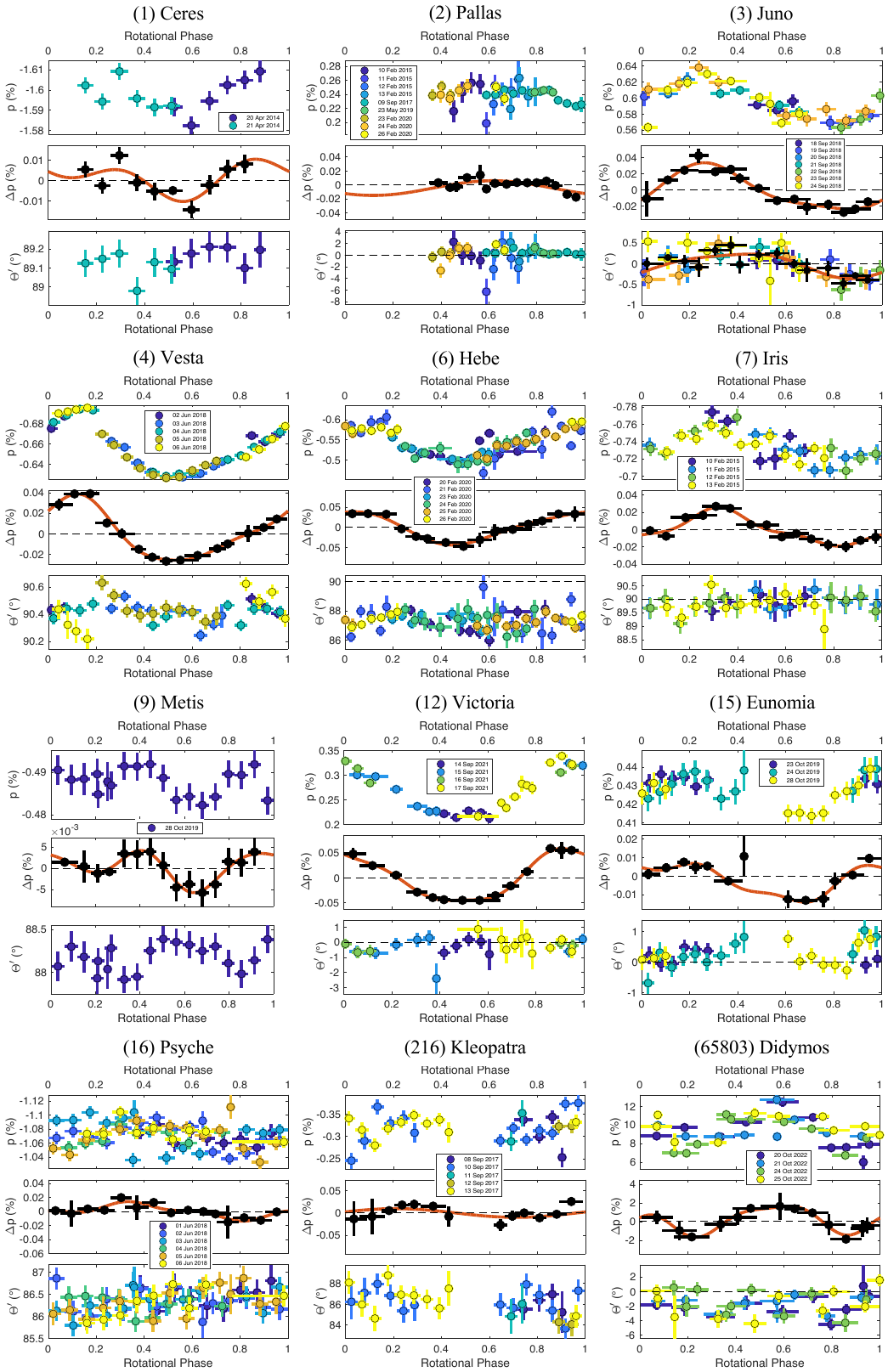}
\caption{Rotation phase-locked measurements of airless bodies obtained with POLISH2 in an unfiltered, 383 to 720 nm bandpass at the Lick 1-m (for 1 Ceres) and Lick 3-m (all other bodies). \textit{Top panels:} Nightly fractional linear polarization $p = P/I$ phased to known rotation period from the JPL Small-Body Database, where rotation phase = 0 is arbitrary. Absolute linear polarization $|p|$ increases toward the top of each panel regardless of whether the object was observed on the positive or negative branch (Figure \ref{phase}). \textit{Center panels:} Phase-locked linear polarimetry binned in rotation phase. Best, third-order Fourier fits are shown as red curves. \textit{Bottom panels:} Linear polarization orientation $\Theta'$ relative to the plane perpendicular to the Sun-body-observer scattering plane. For positive branch (Rayleigh-like) scattering, $\Theta' = 0^\circ$, while $\Theta' = \pm 90^\circ$ for negative branch linear polarization. (3) Juno harbors rotational variations in $\Theta'$ with $> 8 \sigma$ confidence (binned data in black with sinusoidal fit curve in red).}
\label{sample1}
\end{figure*}

Given that phase angle tends to vary by up to $\Delta \alpha \sim 0.4^\circ$ each night due to orbital geometry, linear polarization of a given airless body varies at the detectable level of order $\sim 0.01\%$ each night simply due to geometry (Figure \ref{phasecurve}). To remove the linear polarization variation caused by geometry, we first rotate all Stokes $q,u$ values to the frame perpendicular to the scattering plane (Figure \ref{phasecurve}, blue points). We then identify observations across the run that overlap in rotation phase. Starting with the night with the most data and ending with the least, we interpolate data from pairs of nights at a common set of rotational phases in the overlap region via piecewise cubic Hermite interpolating polynomials (Matlab \texttt{pchip}). Stokes $q'$ and $u'$ offsets for each pair of nights are determined by subtracting interpolated data across the pair of nights and then taking the mean. After this procedure of daisy-chaining overlapping nights of data, phase curve variations caused by geometry are removed from all Stokes $q'$ and $u'$ measurements, which uncovers intrinsic rotational variations (Figure \ref{phasecurve}, red points). We calculate fractional linear polarization $p$ and polarization orientation $\Theta'$ with respect to the plane perpendicular to the scattering plane (Figure \ref{sample1}, top and bottom panels) from daisy-chained Stokes $q'$ and $u'$, and we bin Stokes $q'$, $u'$, $v'$, $p$, and $\Theta'$ in rotation phase (Table \ref{alldatarot} and Figure \ref{sample1}, center panels).

While it is likely that observations of an object obtained years apart will sample different latitudes of the object, we do not obtain sufficient data volume to correct for this. We expect that this effect will tend to dampen the peak-to-peak value of linear polarization variations as different trends are combined. (2) Pallas was observed over the longest time span, and Figure \ref{sample1} shows its polarization variations to be quite weak. However, (2) Pallas observations from 9 Sep 2017 to 26 Feb 2020 are in family, which suggests that smearing of signatures from different latitudes is not a significant effect.

\begin{figure}
\centering
\includegraphics[width=0.47\textwidth]{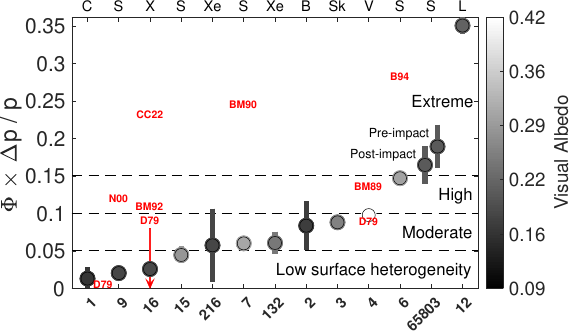}
\caption{Airless bodies with POLISH2 detections of rotational polarization variations sorted by $\Phi \times \Delta p / p$ (normalized peak-to-peak value of rotation phase-locked linear polarization variations) into bins for low, medium, high, and extreme surface heterogeneity. Values of $\Delta p / p$ are multiplied by the phase function $\Phi (\alpha)$ at the time of observation to correct for the shadowed disk present at high phase angles. This reduces $\Delta p / p$ and is only significant for 65803 Didymos at $\alpha \sim 76^\circ$. Relative visual geometric albedo is shown by the greyscale shading of each data point, and taxonomic type is indicated at the top. Both quantities are obtained from the JPL Small-Body Database. No clear correlation between surface heterogeneity $\Delta p / p$ and visual albedo exists. The extreme surface heterogeneity of (12) Victoria is unexpected. Our measurements of (65803) Didymos suggest that surface heterogeneity was higher pre-\text{DART} impact, which implies heterogeneity may be higher for NEOs than for Main Belt Asteroids (see section \ref{sec:didy}). Values and upper limits from the literature are labeled in red \citep{Degewij1979, Broglia1989, Broglia1990, Broglia1992, Broglia1994, Nakayama2000, Castro2022}, where all studies other than \citet[D79]{Degewij1979} report detections with significantly larger variations than we detect in this work.}
\label{dppsort}
\end{figure}

Prior to this investigation, (4) Vesta best displayed repeatable, rotation phase-locked variations of linear polarization \citep{Degewij1979, Broglia1989}, while measurements of (6) Hebe \citep{Broglia1994},  (7) Iris \citep{Broglia1990}, (9) Metis \citep{Nakayama2000}, (16) Psyche \citep{Broglia1992, Castro2022}, (3200) Phaethon \citep{Borisov2018}, and 1998 OR2 \citep{Devogele2024} suggest they may also harbor such variations. However, POLISH2 has not only verified phase-locked variations of linear polarization in (4) Vesta \citep[and Figure \ref{sample1}]{WiktorowiczNofi2015}, but it has also discovered statistically significant rotational variations in nearly every other airless body studied (1 Ceres, 2 Pallas, 3 Juno, 6 Hebe, 7 Iris, 12 Victoria, 15 Eunomia, 16 Psyche, 132 Aethra, 216 Kleopatra, and 65803 Didymos: Figure \ref{sample1}). Rotational variations of (9) Metis in this work are also consistent with a null result since only one night of observations was obtained on this body. Recall that the Umov effect causes linear polarization to be anticorrelated with surface geometric albedo and is independent of shape changes on airless bodies. Therefore, our discoveries of rotation phase-locked linear polarization variations on the above bodies directly translate to estimates of surface albedo and/or compositional heterogeneity. Figures \ref{sample1} through \ref{dppsort} illustrate these detections, where nearly all phase-locked variations are observed to be repeatable across multiple nights.

\startlongtable
\begin{deluxetable*}{cccccccccc}
\tabletypesize{\tiny}
\tablecaption{Rotational Polarimetry of Airless Bodies}
\tablewidth{0pt}
\tablehead{
\colhead{Object} & \colhead{Rot. Phase} & \colhead{$q'$ (\%)}	& \colhead{$u'$ (\%)} & \colhead{$v$ (\%)}	& \colhead{$p$ (\%)}		& \colhead{$\Delta p$ (\%)}	& \colhead{$\Theta'$ ($^\circ$)}}
\startdata
(1) Ceres       & 0.154(34) 	 & $-$1.6014(40)  	 & 0.0490(40)      	 & 0.0195(91)     	 & 1.6022(40)     	 & 0.0053(40)     	 & 89.125(71)    \\
$\cdots$     & 0.225(35) 	 & $-$1.5935(40)  	 & 0.0474(40)      	 & $-$0.0040(92)  	 & 1.5942(40)     	 & $-$0.0027(40)  	 & 89.148(72)    \\
$\cdots$     & 0.296(36) 	 & $-$1.6084(43)  	 & 0.0463(42)      	 & 0.0172(94)     	 & 1.6091(43)     	 & 0.0122(43)     	 & 89.176(75)    \\
$\cdots$     & 0.369(36) 	 & $-$1.5947(42)  	 & 0.0569(41)      	 & 0.0201(94)     	 & 1.5957(42)     	 & $-$0.0011(42)  	 & 88.978(74)    \\
$\cdots$     & 0.441(36) 	 & $-$1.5907(43)  	 & 0.0483(42)      	 & 0.0210(96)     	 & 1.5914(43)     	 & $-$0.0054(43)  	 & 89.130(76)    \\
$\cdots$     & 0.517(40) 	 & $-$1.59101(32) 	 & 0.05101(65)     	 & $-$0.0222(72)  	 & 1.59183(32)    	 & $-$0.00502(32) 	 & 89.082(12)    \\
$\cdots$     & 0.595(35) 	 & $-$1.5817(44)  	 & 0.0490(43)      	 & $-$0.025(10)   	 & 1.5824(44)     	 & $-$0.0144(44)  	 & 89.114(78)    \\
$\cdots$     & 0.670(38) 	 & $-$1.5938(48)  	 & 0.0474(47)      	 & $-$0.027(11)   	 & 1.5945(48)     	 & $-$0.0023(48)  	 & 89.149(84)    \\
$\cdots$     & 0.744(35) 	 & $-$1.6016(46)  	 & 0.0477(45)      	 & 0.004(11)      	 & 1.6023(46)     	 & 0.0055(46)     	 & 89.147(80)    \\
$\cdots$     & 0.816(36) 	 & $-$1.6039(47)  	 & 0.0539(45)      	 & $-$0.003(11)   	 & 1.6048(47)     	 & 0.0079(47)     	 & 89.038(80)    \\
\hline
(2) Pallas      & 0.389(38) 	 & 0.2434(53)     	 & $-$0.0038(49)   	 & 0.0002(50)     	 & $-$0.2434(53)  	 & 0.0030(53)     	 & $-$0.45(57)   \\
$\cdots$     & 0.439(20) 	 & 0.2365(26)     	 & 0.0012(16)      	 & 0.0039(35)     	 & $-$0.2365(26)  	 & $-$0.0039(26)  	 & 0.14(19)      \\
$\cdots$     & 0.464(30) 	 & 0.2375(82)     	 & 0.0118(31)      	 & $-$0.0050(77)  	 & $-$0.2377(81)  	 & $-$0.0027(81)  	 & 1.42(38)      \\
$\cdots$     & 0.509(37) 	 & 0.2507(45)     	 & 0.0059(58)      	 & 0.0158(66)     	 & $-$0.2507(46)  	 & 0.0103(46)     	 & 0.67(66)      \\
$\cdots$     & 0.564(19) 	 & 0.254(15)      	 & $-$0.008(14)    	 & 0.027(13)      	 & $-$0.254(15)   	 & 0.014(15)      	 & $-$0.9(1.6)   \\
$\cdots$     & 0.590(19) 	 & 0.2345(50)     	 & 0.000(12)       	 & $-$0.0127(23)  	 & $-$0.2345(50)  	 & $-$0.0059(50)  	 & 0.0(1.5)      \\
$\cdots$     & 0.625(23) 	 & 0.2424(40)     	 & 0.0044(33)      	 & 0.0100(35)     	 & $-$0.2424(40)  	 & 0.0020(40)     	 & 0.52(39)      \\
$\cdots$     & 0.674(38) 	 & 0.2412(40)     	 & 0.0066(33)      	 & 0.0024(35)     	 & $-$0.2413(40)  	 & 0.0008(40)     	 & 0.79(39)      \\
$\cdots$     & 0.717(31) 	 & 0.2436(47)     	 & 0.0082(38)      	 & $-$0.0082(41)  	 & $-$0.2437(47)  	 & 0.0033(47)     	 & 0.96(45)      \\
$\cdots$     & 0.760(37) 	 & 0.2433(22)     	 & 0.0044(15)      	 & $-$0.0111(30)  	 & $-$0.2434(22)  	 & 0.0030(22)     	 & 0.51(18)      \\
$\cdots$     & 0.799(30) 	 & 0.24335(51)    	 & 0.0057(27)      	 & $-$0.0062(36)  	 & $-$0.24340(52) 	 & 0.00298(52)    	 & 0.68(31)      \\
$\cdots$     & 0.840(30) 	 & 0.24580(19)    	 & 0.00153(78)     	 & $-$0.00179(20) 	 & $-$0.24581(19) 	 & 0.00539(19)    	 & 0.178(91)     \\
$\cdots$     & 0.878(30) 	 & 0.2394(27)     	 & 0.00198(34)     	 & $-$0.0095(27)  	 & $-$0.2395(27)  	 & $-$0.0010(27)  	 & 0.237(41)     \\
$\cdots$     & 0.926(19) 	 & 0.2267(50)     	 & 0.0038(39)      	 & $-$0.0026(42)  	 & $-$0.2267(50)  	 & $-$0.0137(50)  	 & 0.49(49)      \\
$\cdots$     & 0.963(19) 	 & 0.2227(51)     	 & 0.0011(40)      	 & $-$0.0018(43)  	 & $-$0.2227(51)  	 & $-$0.0177(51)  	 & 0.14(51)      \\
\hline
(3) Juno        & 0.023(71) 	 & 0.584(22)      	 & $-$0.0000(32)   	 & 0.0007(24)     	 & $-$0.584(22)   	 & $-$0.011(22)   	 & $-$0.00(16)   \\
$\cdots$     & 0.110(42) 	 & 0.6076(24)     	 & 0.00309(86)     	 & $-$0.0069(31)  	 & $-$0.6076(24)  	 & 0.0121(24)     	 & 0.146(41)     \\
$\cdots$     & 0.179(59) 	 & 0.6197(21)     	 & 0.0012(33)      	 & 0.0008(28)     	 & $-$0.6197(21)  	 & 0.0243(21)     	 & 0.06(15)      \\
$\cdots$     & 0.238(41) 	 & 0.6378(88)     	 & $-$0.0019(53)   	 & 0.0018(35)     	 & $-$0.6378(88)  	 & 0.0424(88)     	 & $-$0.09(24)   \\
$\cdots$     & 0.307(68) 	 & 0.6177(39)     	 & 0.0069(28)      	 & $-$0.0009(25)  	 & $-$0.6178(39)  	 & 0.0224(39)     	 & 0.32(13)      \\
$\cdots$     & 0.372(68) 	 & 0.6208(60)     	 & 0.0095(47)      	 & 0.0129(52)     	 & $-$0.6209(60)  	 & 0.0255(60)     	 & 0.44(22)      \\
$\cdots$     & 0.409(41) 	 & 0.6095(60)     	 & $-$0.0006(47)   	 & 0.0061(51)     	 & $-$0.6095(60)  	 & 0.0141(60)     	 & $-$0.03(22)   \\
$\cdots$     & 0.487(46) 	 & 0.5973(35)     	 & 0.0044(27)      	 & $-$0.0001(30)  	 & $-$0.5973(35)  	 & 0.0019(35)     	 & 0.21(13)      \\
$\cdots$     & 0.563(50) 	 & 0.5819(35)     	 & 0.0051(27)      	 & $-$0.0013(30)  	 & $-$0.5819(35)  	 & $-$0.0135(35)  	 & 0.25(13)      \\
$\cdots$     & 0.635(68) 	 & 0.5839(40)     	 & $-$0.0001(22)   	 & $-$0.0018(27)  	 & $-$0.5839(40)  	 & $-$0.0115(40)  	 & $-$0.00(11)   \\
$\cdots$     & 0.687(41) 	 & 0.5740(96)     	 & $-$0.0033(57)   	 & $-$0.0074(37)  	 & $-$0.5740(96)  	 & $-$0.0215(96)  	 & $-$0.16(28)   \\
$\cdots$     & 0.776(49) 	 & 0.5771(62)     	 & $-$0.0023(37)   	 & 0.0001(22)     	 & $-$0.5771(62)  	 & $-$0.0183(62)  	 & $-$0.11(18)   \\
$\cdots$     & 0.840(51) 	 & 0.5673(43)     	 & $-$0.0096(29)   	 & 0.0028(23)     	 & $-$0.5674(43)  	 & $-$0.0280(43)  	 & $-$0.49(15)   \\
$\cdots$     & 0.889(64) 	 & 0.5716(17)     	 & $-$0.00578(62)  	 & $-$0.00112(88) 	 & $-$0.5717(17)  	 & $-$0.0237(17)  	 & $-$0.290(31)  \\
$\cdots$     & 0.940(49) 	 & 0.5806(28)     	 & $-$0.008110(55) 	 & $-$0.0056(23)  	 & $-$0.5806(28)  	 & $-$0.0148(28)  	 & $-$0.4002(33) \\
\hline
(4) Vesta       & 0.044(59) 	 & $-$0.6830(55)  	 & $-$0.01010(51)  	 & $-$0.0062(80)  	 & 0.6830(55)     	 & 0.0283(55)     	 & 90.424(22)    \\
$\cdots$     & 0.110(44) 	 & $-$0.69337(78) 	 & $-$0.00923(96)  	 & $-$0.0025(75)  	 & 0.69343(78)    	 & 0.03873(78)    	 & 90.381(40)    \\
$\cdots$     & 0.175(39) 	 & $-$0.6936(11)  	 & $-$0.0095(10)   	 & $-$0.00725(53) 	 & 0.6936(11)     	 & 0.0389(11)     	 & 90.392(42)    \\
$\cdots$     & 0.244(47) 	 & $-$0.6649(33)  	 & $-$0.01178(76)  	 & 0.005(12)      	 & 0.6650(33)     	 & 0.0103(33)     	 & 90.508(33)    \\
$\cdots$     & 0.307(55) 	 & $-$0.6545(48)  	 & $-$0.01146(63)  	 & 0.010(11)      	 & 0.6546(48)     	 & $-$0.0001(48)  	 & 90.501(28)    \\
$\cdots$     & 0.378(32) 	 & $-$0.63946(87) 	 & $-$0.00948(67)  	 & 0.007(10)      	 & 0.63953(87)    	 & $-$0.01517(87) 	 & 90.425(30)    \\
$\cdots$     & 0.434(32) 	 & $-$0.63168(71) 	 & $-$0.00837(61)  	 & 0.002(10)      	 & 0.63174(71)    	 & $-$0.02296(71) 	 & 90.379(28)    \\
$\cdots$     & 0.492(27) 	 & $-$0.62778(83) 	 & $-$0.00895(70)  	 & 0.005(14)      	 & 0.62784(83)    	 & $-$0.02686(83) 	 & 90.409(32)    \\
$\cdots$     & 0.551(76) 	 & $-$0.62880(69) 	 & $-$0.009382(59) 	 & 0.001(12)      	 & 0.62887(69)    	 & $-$0.02583(69) 	 & 90.4274(27)   \\
$\cdots$     & 0.623(55) 	 & $-$0.6334(28)  	 & $-$0.00768(54)  	 & 0.006(14)      	 & 0.6334(28)     	 & $-$0.0213(28)  	 & 90.347(25)    \\
$\cdots$     & 0.705(35) 	 & $-$0.64006(90) 	 & $-$0.00781(72)  	 & $-$0.0082(38)  	 & 0.64011(90)    	 & $-$0.01459(90) 	 & 90.350(32)    \\
$\cdots$     & 0.749(27) 	 & $-$0.6446(11)  	 & $-$0.01044(94)  	 & $-$0.00817(93) 	 & 0.6447(11)     	 & $-$0.0100(11)  	 & 90.464(42)    \\
$\cdots$     & 0.834(55) 	 & $-$0.6549(80)  	 & $-$0.0104(25)   	 & $-$0.001(12)   	 & 0.6549(80)     	 & 0.0002(80)     	 & 90.46(11)     \\
$\cdots$     & 0.899(44) 	 & $-$0.6606(37)  	 & $-$0.01051(44)  	 & 0.000(14)      	 & 0.6607(37)     	 & 0.0060(37)     	 & 90.456(19)    \\
$\cdots$     & 0.953(44) 	 & $-$0.6689(26)  	 & $-$0.01045(62)  	 & $-$0.002(14)   	 & 0.6689(26)     	 & 0.0142(26)     	 & 90.448(27)    \\
\hline
(6) Hebe        & 0.031(53) 	 & $-$0.5731(41)  	 & 0.0588(34)      	 & $-$0.0074(37)  	 & 0.5761(41)     	 & 0.0336(41)     	 & 87.07(17)     \\
$\cdots$     & 0.096(41) 	 & $-$0.5738(53)  	 & 0.0541(44)      	 & $-$0.0111(49)  	 & 0.5764(53)     	 & 0.0339(53)     	 & 87.31(22)     \\
$\cdots$     & 0.173(45) 	 & $-$0.5726(40)  	 & 0.0518(33)      	 & $-$0.0001(36)  	 & 0.5749(40)     	 & 0.0324(40)     	 & 87.41(16)     \\
$\cdots$     & 0.236(29) 	 & $-$0.5370(56)  	 & 0.0403(36)      	 & $-$0.0105(50)  	 & 0.5385(56)     	 & $-$0.0040(56)  	 & 87.85(19)     \\
$\cdots$     & 0.296(43) 	 & $-$0.5178(38)  	 & 0.0445(32)      	 & 0.0018(34)     	 & 0.5196(38)     	 & $-$0.0228(38)  	 & 87.55(17)     \\
$\cdots$     & 0.344(32) 	 & $-$0.5117(19)  	 & 0.0514(16)      	 & 0.0078(67)     	 & 0.5143(19)     	 & $-$0.0282(19)  	 & 87.134(87)    \\
$\cdots$     & 0.420(82) 	 & $-$0.5025(47)  	 & 0.0459(39)      	 & 0.0064(41)     	 & 0.5046(47)     	 & $-$0.0379(47)  	 & 87.39(22)     \\
$\cdots$     & 0.494(45) 	 & $-$0.4923(42)  	 & 0.0516(38)      	 & 0.0032(40)     	 & 0.4950(42)     	 & $-$0.0475(42)  	 & 87.01(22)     \\
$\cdots$     & 0.562(51) 	 & $-$0.509(18)   	 & 0.053(13)       	 & 0.0029(31)     	 & 0.511(18)      	 & $-$0.031(18)   	 & 87.05(75)     \\
$\cdots$     & 0.623(52) 	 & $-$0.528(21)   	 & 0.0509(34)      	 & $-$0.003(22)   	 & 0.530(21)      	 & $-$0.013(21)   	 & 87.24(21)     \\
$\cdots$     & 0.700(80) 	 & $-$0.5344(42)  	 & 0.0549(34)      	 & $-$0.0104(37)  	 & 0.5372(41)     	 & $-$0.0053(41)  	 & 87.07(18)     \\
$\cdots$     & 0.765(42) 	 & $-$0.5478(51)  	 & 0.0519(42)      	 & 0.0028(46)     	 & 0.5502(51)     	 & 0.0077(51)     	 & 87.29(22)     \\
$\cdots$     & 0.822(41) 	 & $-$0.5564(53)  	 & 0.0469(43)      	 & 0.0078(47)     	 & 0.5584(53)     	 & 0.0159(53)     	 & 87.59(22)     \\
$\cdots$     & 0.891(47) 	 & $-$0.5730(48)  	 & 0.0545(39)      	 & 0.0046(43)     	 & 0.5756(47)     	 & 0.0331(47)     	 & 87.28(19)     \\
$\cdots$     & 0.959(41) 	 & $-$0.573(18)   	 & 0.055(14)       	 & 0.0050(40)     	 & 0.575(18)      	 & 0.033(18)      	 & 87.25(69)     \\
\hline
(7) Iris        & 0.035(32) 	 & $-$0.7332(21)  	 & 0.008582(86)    	 & 0.0004(32)     	 & 0.7333(21)     	 & $-$0.0013(21)  	 & 89.6647(35)   \\
$\cdots$     & 0.102(34) 	 & $-$0.72688(79) 	 & 0.0040(37)      	 & $-$0.00817(45) 	 & 0.72688(79)    	 & $-$0.00771(79) 	 & 89.84(15)     \\
$\cdots$     & 0.183(62) 	 & $-$0.7482(24)  	 & 0.0164(55)      	 & 0.0024(29)     	 & 0.7484(24)     	 & 0.0138(24)     	 & 89.37(21)     \\
$\cdots$     & 0.256(58) 	 & $-$0.7508(54)  	 & 0.0060(16)      	 & $-$0.0072(52)  	 & 0.7508(54)     	 & 0.0163(54)     	 & 89.772(62)    \\
$\cdots$     & 0.309(58) 	 & $-$0.7613(44)  	 & $-$0.0013(55)   	 & 0.0094(47)     	 & 0.7613(44)     	 & 0.0267(44)     	 & 90.05(21)     \\
$\cdots$     & 0.369(50) 	 & $-$0.7589(46)  	 & 0.0062(25)      	 & 0.0039(49)     	 & 0.7589(46)     	 & 0.0243(46)     	 & 89.766(96)    \\
$\cdots$     & 0.454(62) 	 & $-$0.7403(44)  	 & 0.0006(54)      	 & $-$0.0006(37)  	 & 0.7403(44)     	 & 0.0057(44)     	 & 89.98(21)     \\
$\cdots$     & 0.521(52) 	 & $-$0.7400(50)  	 & 0.0034(56)      	 & 0.0094(53)     	 & 0.7400(50)     	 & 0.0054(50)     	 & 89.87(22)     \\
$\cdots$     & 0.584(52) 	 & $-$0.7260(42)  	 & 0.0024(60)      	 & $-$0.0069(56)  	 & 0.7260(42)     	 & $-$0.0086(42)  	 & 89.90(23)     \\
$\cdots$     & 0.645(52) 	 & $-$0.7299(46)  	 & 0.0036(26)      	 & 0.0091(49)     	 & 0.7299(46)     	 & $-$0.0047(46)  	 & 89.86(10)     \\
$\cdots$     & 0.707(52) 	 & $-$0.7237(46)  	 & $-$0.0011(58)   	 & 0.0084(49)     	 & 0.7237(46)     	 & $-$0.0109(46)  	 & 90.04(23)     \\
$\cdots$     & 0.778(35) 	 & $-$0.7169(64)  	 & 0.0041(79)      	 & $-$0.0046(68)  	 & 0.7169(64)     	 & $-$0.0177(64)  	 & 89.84(31)     \\
$\cdots$     & 0.851(32) 	 & $-$0.7149(64)  	 & 0.0026(13)      	 & $-$0.0109(69)  	 & 0.7149(64)     	 & $-$0.0197(64)  	 & 89.898(52)    \\
$\cdots$     & 0.912(32) 	 & $-$0.7220(25)  	 & $-$0.00237(63)  	 & 0.0008(65)     	 & 0.7220(25)     	 & $-$0.0126(25)  	 & 90.094(25)    \\
$\cdots$     & 0.975(31) 	 & $-$0.7257(79)  	 & 0.0116(98)      	 & $-$0.0056(84)  	 & 0.7257(79)     	 & $-$0.0089(79)  	 & 89.54(39)     \\
\hline
(9) Metis       & 0.062(59) 	 & $-$0.4884(10)  	 & 0.0311(20)      	 & $-$0.0011(13)  	 & 0.4894(10)     	 & 0.0014(10)     	 & 88.18(12)     \\
$\cdots$     & 0.149(29) 	 & $-$0.4874(38)  	 & 0.0310(32)      	 & 0.0069(35)     	 & 0.4884(38)     	 & 0.0004(38)     	 & 88.18(19)     \\
$\cdots$     & 0.210(31) 	 & $-$0.4856(22)  	 & 0.0332(18)      	 & $-$0.0022(11)  	 & 0.4868(22)     	 & $-$0.0012(22)  	 & 88.05(11)     \\
$\cdots$     & 0.263(31) 	 & $-$0.48621(30) 	 & 0.0305(20)      	 & 0.0015(21)     	 & 0.48717(32)    	 & $-$0.00083(32) 	 & 88.20(12)     \\
$\cdots$     & 0.329(29) 	 & $-$0.4901(33)  	 & 0.0358(27)      	 & $-$0.0004(30)  	 & 0.4914(33)     	 & 0.0034(33)     	 & 87.91(16)     \\
$\cdots$     & 0.388(29) 	 & $-$0.4901(33)  	 & 0.0352(27)      	 & 0.0001(31)     	 & 0.4914(33)     	 & 0.0034(33)     	 & 87.95(16)     \\
$\cdots$     & 0.446(29) 	 & $-$0.4910(32)  	 & 0.0300(27)      	 & 0.0047(30)     	 & 0.4919(32)     	 & 0.0039(32)     	 & 88.25(16)     \\
$\cdots$     & 0.505(29) 	 & $-$0.4879(32)  	 & 0.0275(26)      	 & $-$0.0015(29)  	 & 0.4887(32)     	 & 0.0007(32)     	 & 88.39(16)     \\
$\cdots$     & 0.563(29) 	 & $-$0.4827(32)  	 & 0.0278(26)      	 & 0.0001(30)     	 & 0.4835(32)     	 & $-$0.0045(32)  	 & 88.35(16)     \\
$\cdots$     & 0.622(29) 	 & $-$0.4834(32)  	 & 0.0284(26)      	 & $-$0.0027(29)  	 & 0.4843(32)     	 & $-$0.0037(32)  	 & 88.32(16)     \\
$\cdots$     & 0.680(29) 	 & $-$0.4814(32)  	 & 0.0295(27)      	 & 0.0014(30)     	 & 0.4823(32)     	 & $-$0.0057(32)  	 & 88.25(16)     \\
$\cdots$     & 0.739(29) 	 & $-$0.4833(32)  	 & 0.0287(27)      	 & $-$0.0018(29)  	 & 0.4842(32)     	 & $-$0.0038(32)  	 & 88.30(16)     \\
$\cdots$     & 0.798(29) 	 & $-$0.4885(32)  	 & 0.0324(27)      	 & $-$0.0015(29)  	 & 0.4896(32)     	 & 0.0016(32)     	 & 88.10(16)     \\
$\cdots$     & 0.856(29) 	 & $-$0.4881(33)  	 & 0.0345(28)      	 & $-$0.0010(30)  	 & 0.4893(33)     	 & 0.0013(33)     	 & 87.98(16)     \\
$\cdots$     & 0.915(29) 	 & $-$0.4908(33)  	 & 0.0319(28)      	 & $-$0.0024(31)  	 & 0.4918(33)     	 & 0.0038(33)     	 & 88.14(16)     \\
\hline
(12) Victoria   & 0.040(53) 	 & 0.314(12)      	 & $-$0.0051(23)   	 & $-$0.0030(26)  	 & $-$0.314(12)   	 & 0.048(12)      	 & $-$0.46(21)   \\
$\cdots$     & 0.120(53) 	 & 0.2901(38)     	 & $-$0.00672(75)  	 & 0.0013(28)     	 & $-$0.2902(38)  	 & 0.0244(38)     	 & $-$0.664(75)  \\
$\cdots$     & 0.218(26) 	 & 0.2708(53)     	 & $-$0.0016(40)   	 & 0.0053(45)     	 & $-$0.2708(53)  	 & 0.0050(53)     	 & $-$0.16(42)   \\
$\cdots$     & 0.306(44) 	 & 0.2364(52)     	 & 0.0012(40)      	 & $-$0.0022(44)  	 & $-$0.2364(52)  	 & $-$0.0295(52)  	 & 0.15(48)      \\
$\cdots$     & 0.371(30) 	 & 0.22540(23)    	 & $-$0.0021(36)   	 & 0.0040(41)     	 & $-$0.22540(23) 	 & $-$0.04044(23) 	 & $-$0.27(46)   \\
$\cdots$     & 0.417(26) 	 & 0.2211(47)     	 & $-$0.0055(36)   	 & $-$0.0010(40)  	 & $-$0.2212(47)  	 & $-$0.0447(47)  	 & $-$0.71(47)   \\
$\cdots$     & 0.494(51) 	 & 0.2199(34)     	 & $-$0.0002(16)   	 & $-$0.0010(12)  	 & $-$0.2199(34)  	 & $-$0.0459(34)  	 & $-$0.03(21)   \\
$\cdots$     & 0.565(86) 	 & 0.2199(37)     	 & 0.0029(29)      	 & 0.0019(33)     	 & $-$0.2199(37)  	 & $-$0.0460(37)  	 & 0.38(38)      \\
$\cdots$     & 0.631(34) 	 & 0.2294(69)     	 & $-$0.0020(37)   	 & $-$0.0198(59)  	 & $-$0.2294(69)  	 & $-$0.0364(69)  	 & $-$0.25(46)   \\
$\cdots$     & 0.695(36) 	 & 0.2491(49)     	 & $-$0.00257(96)  	 & 0.0102(42)     	 & $-$0.2491(49)  	 & $-$0.0167(49)  	 & $-$0.30(11)   \\
$\cdots$     & 0.763(26) 	 & 0.2781(28)     	 & 0.0010(41)      	 & 0.0070(50)     	 & $-$0.2781(28)  	 & 0.0123(28)     	 & 0.11(42)      \\
$\cdots$     & 0.860(22) 	 & 0.3250(58)     	 & $-$0.0043(45)   	 & 0.0042(51)     	 & $-$0.3250(58)  	 & 0.0592(58)     	 & $-$0.38(40)   \\
$\cdots$     & 0.906(28) 	 & 0.321(17)      	 & 0.0006(14)      	 & $-$0.0024(26)  	 & $-$0.321(17)   	 & 0.055(17)      	 & 0.05(12)      \\
$\cdots$     & 0.948(31) 	 & 0.3211(21)     	 & $-$0.0052(22)   	 & $-$0.0046(29)  	 & $-$0.3212(21)  	 & 0.0553(21)     	 & $-$0.47(20)   \\
\hline
(15) Eunomia    & 0.028(49) 	 & 0.4286(29)     	 & $-$0.0006(24)   	 & $-$0.0012(84)  	 & $-$0.4286(29)  	 & 0.0010(29)     	 & $-$0.04(16)   \\
$\cdots$     & 0.102(51) 	 & 0.4320(24)     	 & 0.0003(20)      	 & 0.0002(16)     	 & $-$0.4320(24)  	 & 0.0044(24)     	 & 0.02(13)      \\
$\cdots$     & 0.176(27) 	 & 0.43504(88)    	 & 0.0052(24)      	 & $-$0.00420(56) 	 & $-$0.43506(88) 	 & 0.00746(88)    	 & 0.34(16)      \\
$\cdots$     & 0.225(27) 	 & 0.4323(35)     	 & 0.0052(12)      	 & 0.0039(14)     	 & $-$0.4323(35)  	 & 0.0047(35)     	 & 0.346(78)     \\
$\cdots$     & 0.274(27) 	 & 0.43295(21)    	 & 0.0031(28)      	 & $-$0.0024(22)  	 & $-$0.43296(21) 	 & 0.00536(21)    	 & 0.21(18)      \\
$\cdots$     & 0.361(59) 	 & 0.4248(19)     	 & 0.0058(30)      	 & $-$0.0000(20)  	 & $-$0.4249(19)  	 & $-$0.0027(19)  	 & 0.39(20)      \\
$\cdots$     & 0.427(12) 	 & 0.438(11)      	 & 0.0125(81)      	 & 0.0068(46)     	 & $-$0.438(11)   	 & 0.011(11)      	 & 0.82(53)      \\
$\cdots$     & 0.611(24) 	 & 0.4150(47)     	 & 0.0110(42)      	 & $-$0.0020(45)  	 & $-$0.4152(47)  	 & $-$0.0124(47)  	 & 0.76(29)      \\
$\cdots$     & 0.684(49) 	 & 0.41449(82)    	 & 0.0016(13)      	 & $-$0.0033(29)  	 & $-$0.41449(82) 	 & $-$0.01310(82) 	 & 0.108(91)     \\
$\cdots$     & 0.757(24) 	 & 0.4152(44)     	 & $-$0.0014(39)   	 & 0.0029(41)     	 & $-$0.4152(44)  	 & $-$0.0124(44)  	 & $-$0.10(27)   \\
$\cdots$     & 0.806(24) 	 & 0.4249(45)     	 & $-$0.0016(41)   	 & $-$0.0058(43)  	 & $-$0.4249(45)  	 & $-$0.0027(45)  	 & $-$0.11(27)   \\
$\cdots$     & 0.880(49) 	 & 0.4281(14)     	 & 0.0026(27)      	 & 0.0034(22)     	 & $-$0.4281(14)  	 & 0.0005(14)     	 & 0.17(18)      \\
$\cdots$     & 0.948(49) 	 & 0.4369(24)     	 & 0.0086(25)      	 & 0.0018(18)     	 & $-$0.4370(24)  	 & 0.0094(24)     	 & 0.56(16)      \\
\hline
(16) Psyche     & 0.030(42) 	 & $-$1.0624(51)  	 & 0.1368(53)      	 & $-$0.0023(51)  	 & 1.0711(51)     	 & 0.0011(51)     	 & 86.33(14)     \\
$\cdots$     & 0.096(44) 	 & $-$1.057(18)   	 & 0.1447(47)      	 & $-$0.007(18)   	 & 1.067(18)      	 & $-$0.003(18)   	 & 86.10(14)     \\
$\cdots$     & 0.165(44) 	 & $-$1.0636(40)  	 & 0.1461(41)      	 & $-$0.008(20)   	 & 1.0736(40)     	 & 0.0036(40)     	 & 86.09(11)     \\
$\cdots$     & 0.235(44) 	 & $-$1.0624(41)  	 & 0.1426(42)      	 & $-$0.0060(40)  	 & 1.0719(41)     	 & 0.0018(41)     	 & 86.18(11)     \\
$\cdots$     & 0.304(44) 	 & $-$1.0802(41)  	 & 0.1438(43)      	 & $-$0.0099(41)  	 & 1.0897(41)     	 & 0.0197(41)     	 & 86.21(11)     \\
$\cdots$     & 0.369(50) 	 & $-$1.067(18)   	 & 0.1353(39)      	 & 0.003(29)      	 & 1.076(17)      	 & 0.006(17)      	 & 86.39(12)     \\
$\cdots$     & 0.440(49) 	 & $-$1.0732(38)  	 & 0.1443(40)      	 & 0.004(28)      	 & 1.0829(38)     	 & 0.0128(38)     	 & 86.17(10)     \\
$\cdots$     & 0.512(51) 	 & $-$1.0601(39)  	 & 0.1304(41)      	 & 0.002(24)      	 & 1.0681(39)     	 & $-$0.0020(39)  	 & 86.49(11)     \\
$\cdots$     & 0.583(50) 	 & $-$1.0625(38)  	 & 0.1406(39)      	 & $-$0.005(29)   	 & 1.0718(38)     	 & 0.0017(38)     	 & 86.23(10)     \\
$\cdots$     & 0.647(48) 	 & $-$1.0623(42)  	 & 0.1276(43)      	 & $-$0.0162(42)  	 & 1.0699(42)     	 & $-$0.0002(42)  	 & 86.57(12)     \\
$\cdots$     & 0.698(67) 	 & $-$1.0579(46)  	 & 0.1347(47)      	 & 0.005(25)      	 & 1.0664(46)     	 & $-$0.0037(46)  	 & 86.37(13)     \\
$\cdots$     & 0.749(47) 	 & $-$1.046(24)   	 & 0.1356(43)      	 & 0.017(26)      	 & 1.055(24)      	 & $-$0.015(24)   	 & 86.31(14)     \\
$\cdots$     & 0.817(47) 	 & $-$1.0534(47)  	 & 0.1273(49)      	 & 0.008(18)      	 & 1.0611(47)     	 & $-$0.0090(47)  	 & 86.55(13)     \\
$\cdots$     & 0.886(47) 	 & $-$1.0484(47)  	 & 0.1385(49)      	 & 0.002(22)      	 & 1.0575(47)     	 & $-$0.0126(47)  	 & 86.24(13)     \\
$\cdots$     & 0.953(47) 	 & $-$1.0612(49)  	 & 0.1321(51)      	 & 0.013(26)      	 & 1.0693(49)     	 & $-$0.0007(49)  	 & 86.45(14)     \\
\hline
(216) Kleopatra & 0.037(54) 	 & $-$0.302(39)   	 & 0.0333(78)      	 & 0.0154(88)     	 & 0.304(39)      	 & $-$0.014(39)   	 & 86.85(84)     \\
$\cdots$     & 0.113(54) 	 & $-$0.307(40)   	 & 0.0374(75)      	 & $-$0.0063(45)  	 & 0.309(39)      	 & $-$0.009(39)   	 & 86.52(83)     \\
$\cdots$     & 0.183(32) 	 & $-$0.3214(49)  	 & 0.0240(83)      	 & 0.0089(28)     	 & 0.3222(49)     	 & 0.0048(49)     	 & 87.86(74)     \\
$\cdots$     & 0.237(32) 	 & $-$0.3320(31)  	 & 0.0433(82)      	 & $-$0.0022(42)  	 & 0.3347(33)     	 & 0.0173(33)     	 & 86.28(70)     \\
$\cdots$     & 0.288(27) 	 & $-$0.334(11)   	 & 0.0398(26)      	 & $-$0.0007(13)  	 & 0.336(11)      	 & 0.019(11)      	 & 86.60(25)     \\
$\cdots$     & 0.367(52) 	 & $-$0.3291(35)  	 & 0.0452(56)      	 & $-$0.0029(48)  	 & 0.3321(36)     	 & 0.0147(36)     	 & 86.09(48)     \\
$\cdots$     & 0.432(14) 	 & $-$0.308(23)   	 & 0.027(18)       	 & $-$0.029(21)   	 & 0.309(23)      	 & $-$0.008(23)   	 & 87.5(1.7)     \\
$\cdots$     & 0.646(27) 	 & $-$0.287(14)   	 & 0.041(11)       	 & $-$0.001(12)   	 & 0.290(14)      	 & $-$0.027(14)   	 & 86.0(1.1)     \\
$\cdots$     & 0.696(32) 	 & $-$0.306(12)   	 & 0.0493(16)      	 & $-$0.0273(84)  	 & 0.310(11)      	 & $-$0.007(11)   	 & 85.42(23)     \\
$\cdots$     & 0.745(39) 	 & $-$0.3149(97)  	 & 0.0374(75)      	 & 0.0113(84)     	 & 0.3170(97)     	 & $-$0.0004(97)  	 & 86.61(68)     \\
$\cdots$     & 0.809(27) 	 & $-$0.3039(75)  	 & 0.0358(18)      	 & 0.0032(86)     	 & 0.3060(74)     	 & $-$0.0114(74)  	 & 86.64(19)     \\
$\cdots$     & 0.880(41) 	 & $-$0.3104(81)  	 & 0.0484(63)      	 & 0.014(31)      	 & 0.3141(81)     	 & $-$0.0033(81)  	 & 85.57(58)     \\
$\cdots$     & 0.941(49) 	 & $-$0.3354(93)  	 & 0.0692(73)      	 & 0.0164(82)     	 & 0.3424(92)     	 & 0.0250(92)     	 & 84.17(62)     \\
\hline
(65803) Didymos & 0.073(66) 	 & 9.62(71)       	 & $-$0.39(14)     	 & $-$0.21(56)    	 & $-$9.62(71)    	 & 0.45(71)       	 & $-$1.17(42)   \\
$\cdots$     & 0.166(74) 	 & 8.2(1.0)       	 & $-$0.27(16)     	 & 0.04(16)       	 & $-$8.2(1.0)    	 & $-$1.0(1.0)    	 & $-$0.95(56)   \\
$\cdots$     & 0.217(74) 	 & 7.52(26)       	 & $-$0.15(22)     	 & 0.41(75)       	 & $-$7.52(26)    	 & $-$1.64(26)    	 & $-$0.59(82)   \\
$\cdots$     & 0.338(67) 	 & 8.82(79)       	 & $-$1.029(88)    	 & $-$0.03(66)    	 & $-$8.88(79)    	 & $-$0.29(79)    	 & $-$3.32(41)   \\
$\cdots$     & 0.409(79) 	 & 9.60(83)       	 & $-$0.61(12)     	 & 0.27(22)       	 & $-$9.62(83)    	 & 0.45(83)       	 & $-$1.82(39)   \\
$\cdots$     & 0.462(75) 	 & 10.57(26)      	 & $-$0.53(22)     	 & $-$0.1(1.4)    	 & $-$10.59(26)   	 & 1.42(26)       	 & $-$1.43(59)   \\
$\cdots$     & 0.584(81) 	 & 10.8(1.5)      	 & $-$0.83(16)     	 & 0.3(1.1)       	 & $-$10.8(1.5)   	 & 1.6(1.5)       	 & $-$2.20(53)   \\
$\cdots$     & 0.645(41) 	 & 10.54(31)      	 & $-$0.17(26)     	 & $-$0.78(27)    	 & $-$10.54(31)   	 & 1.37(31)       	 & $-$0.47(70)   \\
$\cdots$     & 0.719(66) 	 & 10.10(28)      	 & $-$0.178(83)    	 & $-$1.09(14)    	 & $-$10.10(28)   	 & 0.93(28)       	 & $-$0.50(24)   \\
$\cdots$     & 0.787(70) 	 & 8.8(1.1)       	 & $-$0.84(13)     	 & $-$0.05(70)    	 & $-$8.8(1.1)    	 & $-$0.3(1.1)    	 & $-$2.75(56)   \\
$\cdots$     & 0.858(66) 	 & 7.24(20)       	 & $-$0.78(16)     	 & 0.24(17)       	 & $-$7.28(20)    	 & $-$1.88(20)    	 & $-$3.06(64)   \\
$\cdots$     & 0.928(50) 	 & 8.44(90)       	 & $-$0.21(13)     	 & 0.84(21)       	 & $-$8.44(90)    	 & $-$0.73(90)    	 & $-$0.71(45)   \\
$\cdots$     & 0.946(66) 	 & 8.74(89)       	 & $-$0.42(17)     	 & 0.11(85)       	 & $-$8.75(89)    	 & $-$0.42(89)    	 & $-$1.37(58)     
\label{alldatarot}
\enddata
\end{deluxetable*}

 Peak-to-peak linear polarization variations $\Delta p$ are calculated either from Fourier fits to binned linear polarization $p$ as a function of rotational phase $\psi$ (Equation \ref{fouriereq}) or the difference between maximum and minimum binned linear polarization values (Figure \ref{sample1}, center panels), whichever provides the lowest uncertainty:
 
\begin{eqnarray}
\label{fouriereq}
p(\psi) & = &  a_0 + a_1 \cos(\psi) + b_1 \sin(\psi) + a_2 \cos(2\psi) \\
\nonumber & + & b_2 \sin(2\psi) + a_3 \cos(3\psi) + b_3 \sin(3\psi).
\end{eqnarray}
 
 \noindent Fourier fits of first through third order are performed, and fits are discarded if their extrema lie outside the range of measurements. This prevents oscillatory fits to sparse datasets such as (2) Pallas, as $\Delta p$ must be based on data and not on wild interpolation. The best fit for each object (first through third order Fourier fit) is chosen to be the fit order with the highest adjusted coefficient of determination $R^2$. If obtained from Fourier fits, $\Delta p$ is given by the peak-to-peak value of the best Fourier fit (red curves in Figure \ref{sample1}, center panels). Best-fit Fourier coefficients for each object are listed in Table \ref{fourier} along with adjusted $R^2$ of the best Fourier fit, rotation period, and the epoch of zero arbitrary rotation phase $T_0$ in MJD. While Fourier fits to (2) Pallas and (216) Kleopatra have low adjusted $R^2$ values, they are sufficient to estimate peak-to-peak linear polarization variations $\Delta p$ of the objects.
 
 Figure \ref{dppsort} compares phase-locked linear polarization variations $\Delta p$ across our sample to estimates from the literature. (4) Vesta harbors some of the strongest variations, which suggests why it enabled the first and most statistically significant detection of this effect \citep{Degewij1979}. However, POLISH2 shows that (4) Vesta shares this distinction with (6) Hebe, (12) Victoria, and (65803) Didymos. We find that the peak-to-peak value of linear polarization variations $\Delta p$ is correlated with fractional linear polarization $p$, which is dominated by Sun-body-observer phase angle $\alpha$ (Figure \ref{phase}). As Figure \ref{phase} shows, airless bodies tend to be strongly polarized for $\alpha > 30^\circ$, which suggests their rotational polarization variations $\Delta p$ will be pronounced at high $\alpha$. We have confirmed this via observations of (65803) Didymos at $\alpha \sim 76^\circ$, whose peak-to-peak value of polarization variations ($\Delta p/p \sim 3\%$) is two orders of magnitude larger than those of the MBAs in Figure \ref{sample1} ($\Delta p/p \sim 0.04\%$).
 
 Additionally, even though our observations of (4) Vesta were obtained $\sim 40$ years after the discovery of its polarization variations \citep{Degewij1979}, we measure the same value of $\Delta p / p \sim 0.1$ as in that paper. Thus, normalization of $\Delta p$ by $p$ enables intrinsic surface heterogeneity to be compared between one airless body and another (Figure \ref{dppsort}) and is relatively tolerant of changes in sub-observer latitude. Indeed, the quantity $\Delta p / p$ highlights (4) Vesta's large surface heterogeneity and (1) Ceres' relatively featureless surface as expected. It also exposes the perhaps unexpectedly low heterogeneity of (9) Metis and (16) Psyche as well as the high heterogeneity of (6) Hebe, (12) Victoria, and (65803) Didymos.
 
 Figure \ref{dppsort} also shows that surface heterogeneity $\Delta p / p$ is not correlated with disk-integrated visual geometric albedo from the JPL Small-Body Database. To be clear, peak-to-peak polarization variation $\Delta p / p$ is indeed correlated with peak-to-peak surface albedo variation, but it is not correlated with disk-integrated albedo. Thus, the mean, intrinsic reflectivity of an airless body appears to have no bearing on its variation in surface albedo. In Figure \ref{dppsort}, $\Delta p / p$ is multiplied by the phase function $\Phi(\alpha)$ at the mean phase angle of observation to correct for the fraction of the shadowed disk present at high $\alpha$ \citep{Russell1916}:

\begin{deluxetable*}{cccccccccc}
\tabletypesize{\tiny}
\tablecaption{Fourier Fits to Rotational Linear Polarization}
\tablewidth{0pt}
\tablehead{
\colhead{Object} & \colhead{$a_1$ (\%)} & \colhead{$b_1$ (\%)} & \colhead{$a_2$ (\%)} & \colhead{$b_2$ (\%)} & \colhead{$a_3$ (\%)} & \colhead{$b_3$ (\%)} & \colhead{Adj. $R^2$} & \colhead{Rot. Per. (h)}	& \colhead{$T_0$ (MJD)} }
\startdata
(1) Ceres       & 0.0065(67)    	 & 0.0002(30)    	 & $-$0.0034(52) 	 & $-$0.0046(34)  	 & $-$       	 		& $-$      			&  0.319		& 9.07	& 61041.06179  \\
(2) Pallas      & $-$0.0078(32) 	 & $-$0.0070(45) 	 & $-$      	 		& $-$       	 		& $-$       	 		& $-$      			&  0.153		& 7.81	& 61041.28875  \\
(3) Juno        & $-$0.0052(24) 	 & 0.0270(23)    	 & $-$0.0069(25) 	 & 0.0007(23)     	 & $-$       	 		& $-$      			&  0.920		& 7.21	& 61041.02875  \\
(4) Vesta       & 0.0262(13)    	 & 0.0134(13)    	 & $-$0.0025(13) 	 & 0.0072(13)     	 & $-$0.0024(13)  	 & 0.0013(12)   		&  0.979		& 5.34	& 61041.11219  \\
(6) Hebe        & 0.0392(26)    	 & $-$0.0047(25) 	 & $-$0.0018(25) 	 & 0.0066(26)     	 & $-$       	 		& $-$      			&  0.949		& 7.27	& 61041.05419  \\
(7) Iris        & $-$0.0081(19) 	 & 0.0167(19)    	 & $-$0.0020(19) 	 & $-$0.0026(20)  	 & 0.0038(19)     	 & $-$0.0010(20)	&  0.886		& 7.14	& 61041.22888  \\
(9) Metis       & 0.00141(61)   	 & 0.00167(51)   	 & 0.00165(59)   	 & $-$0.00300(52) 	 & $-$0.00027(58) 	 & 0.00075(52) 		&  0.820		& 5.08	& 61041.00037  \\
(12) Victoria   & 0.0521(16)    	 & $-$0.0102(17) 	 & 0.0003(17)    	 & $-$0.0069(16)  	 & $-$0.0057(18)  	 & $-$0.0018(16)	&  0.991		& 8.66	& 61041.04256  \\
(15) Eunomia    & 0.0076(29)    	 & 0.0061(14)    	 & 0.0002(25)    	 & $-$0.0018(15)  	 & $-$0.0008(23)  	 & $-$0.0025(15)	&  0.830		& 6.08	& 61041.00979  \\
(16) Psyche     & $-$0.0047(24) & 0.0086(23)    	 & 0.0012(24)    	 & $-$0.0008(23)  	 & 0.0039(24)     	 & $-$0.0010(23)	&  0.543		& 4.20	& 61041.13233  \\
(216) Kleopatra & 0.0023(78)     & 0.0094(61)    	 & $-$      	 		& $-$       	 		& $-$       	 		& $-$      			& $-$0.142	& 5.38	& 61041.14062  \\
(65803) Didymos & $-$0.86(16)    & $-$0.49(15)   	 & 0.79(16)      	 	& 0.65(14)       	 	& 0.39(16)       	 	& 0.44(14)        		&  0.913		& 2.26	& 61041.06903
\label{fourier}
\enddata
\tablecomments{Coefficients $a_1$, $b_1$, $a_2$, $b_2$, $a_3$, $b_3$, and adjusted $R^2$ are for linear polarization fits to Equation \ref{fouriereq} for each object.}
\end{deluxetable*}
 
\begin{equation}
\Phi(\alpha) =  [\sin{\alpha} + (\pi - \alpha) \cos{\alpha}] / \pi.
\end{equation}

\noindent This is because high $\alpha$ observations of (65803) Didymos, for example, would otherwise be biased toward large values of $\Delta p / p$ since a relatively small, polarized spot fills a greater fraction of the lit disk at high $\alpha$. Correction by the phase function acts to reduce measured $\Delta p / p$, since $\Phi(\alpha) \leq 1$. Thus, we attempt to be conservative in the estimate of spatial heterogeneity. Table \ref{dpp} lists relative intrinsic linear polarization variability ($\Phi \times \Delta p/p$) for the objects in our sample. Note that (65803) Didymos is listed twice in an attempt to estimate intrinsic variations pre- and post-\textit{DART} impact (section \ref{sec:didy}). Also listed in Table \ref{dpp} are visual albedo from the JPL Small-Body Database; the confidence of rejecting a null, constant polarization model from a $\chi^2$ test (in standard deviation units $\sigma$), and reduced $\chi^2$. Thus, of the objects in Figure \ref{sample1}, the rotational variability of only (9) Metis may be a false positive caused by measurement uncertainty.

While our observations of (9) Metis are of insufficient quality to reject the null hypothesis of a constant polarization model, its measured intrinsic $\Phi \times \Delta p/p$ (Table \ref{dpp}) is an order of magnitude smaller than the value of $\Phi \times \Delta p/p = 0.99 \times 0.1\% / 0.85\% \sim 0.12$ from \citet{Nakayama2000} data. Our estimate of $\Phi \times \Delta p/p = 0.0203 \pm 0.0032$ rejects that value of $\Phi \times \Delta p/p \sim 0.12$ with $30 \sigma$ confidence. Indeed, we measure (9) Metis' value of $\Phi \times \Delta p/p$ to be in the 15th percentile in our sample, while replacing it with that from \citet{Nakayama2000} would place it in the 69th percentile.

To our knowledge, we find the same issue with all other prior reports in the literature \citep{Broglia1989, Broglia1990, Broglia1992, Broglia1994, Castro2022} except \citet{Degewij1979}, where reported polarization variations are significantly larger than those measured in this work. We note that while this body of literature reports $\Phi \times \Delta p/p > 0.1$ for all prior detections (Figure \ref{dppsort}), our work shows a smooth spectrum of variations where 10 out of 13 detections have $\Phi \times \Delta p/p < 0.1$. For (6) Hebe, \citet{Broglia1994} also report sinusoidal variations in $\Theta$ with amplitude $\Delta\Theta = 4\, \fdg 20 \pm 1\, \fdg 56$, while our measurements suggest this to be at most $\Delta\Theta = 0.31^\circ$ with a detection of only $1.7\sigma$ confidence.

The rotational variations of (16) Psyche were verified across four nights, and measured $\Phi(17^\circ) \times \Delta p/p = 0.96 \times 0.14 / 0.58 = 0.23$ \citep{Castro2022}. This value contrasts with the \citet{Broglia1992} estimate of $\Phi(8^\circ) \times \Delta p/p = 0.99 \times 0.11 = 0.11$, the \citet{Degewij1979} upper limit of $\Phi(5\, \fdg 7) \times \Delta p/p < 0.1$, and our estimate of $\Phi \times \Delta p/p= 0.0257 \pm 0.0062$ (Figure \ref{dppsort}). Strikingly, of all the prior detections in Figure \ref{dppsort} (\citealt[1 Ceres, 4 Vesta, and 16 Psyche]{Degewij1979}; \citealt[4 Vesta]{Broglia1989}; \citealt[7 Iris]{Broglia1990}; \citealt[16 Psyche]{Broglia1992, Castro2022}; \citealt[6 Hebe]{Broglia1994}; \citealt[9 Metis]{Nakayama2000}), measurements of all three asteroids from \citet{Degewij1979}, and only those three, are consistent with our measurements.

\begin{deluxetable}{cccccccccc}
\tabletypesize{\tiny}
\tablecaption{Relative Intrinsic Linear Polarization Variability}
\tablewidth{0pt}
\tablehead{
\colhead{Object} & \colhead{Visual Albedo} & \colhead{$\Phi \times \Delta p/p$}	& \colhead{Null Rej. ($\sigma$)}	& \colhead{$\chi^2 / \nu$}}
\startdata
(1) Ceres       & 0.0900(30) 	 & 0.0166(38) 	 & 4.5 	 & 4.5   \\
(2) Pallas      & 0.1010(80) 	 & 0.084(33)  	 & 6.7 	 & 5.7   \\
(3) Juno        & 0.214(26)  	 	& 0.0880(76) 	 & $> 8$ 	 & 38.7  \\
(4) Vesta       & 0.423(53)  	 & 0.0996(21) 	 & $> 8$ 	 & 525 \\
(6) Hebe        & 0.2679(80) 	 & 0.1469(96) 	 & $> 8$ 	 & 46.0  \\
(7) Iris        & 0.277(30)  	 	& 0.0599(63) 	 & $> 8$ 	 & 15.2  \\
(9) Metis       & 0.118      	 	& 0.0203(32) 	 & 1.3 	 & 1.3   \\
(12) Victoria   & 0.163(27)  	 & 0.351(11)  	 & $> 8$ 	 & 240 \\
(15) Eunomia    & 0.248(42)  	 & 0.044(11)  	 & $> 8$ 	 & 45.2  \\
(16) Psyche     & 0.1203(40) 	 & 0.0257(62) 	 & 4.2 	 & 3.3   \\
(132) Aethra    & 0.199(15)  	 & 0.070(11)  	 & $> 8$ 	 & 15.0  \\
(216) Kleopatra & 0.1164(40) 	 & 0.057(49)  	 & 3.4 	 & 2.8   \\
(65803) Didymos pre & 0.150(40)  	 & 0.190(29)  	 & $> 8$ 	 & 15.7  \\
(65803) Didymos post & $\cdots$  	 & 0.165(25)  	 & $> 8$ 	 & 15.7
\label{dpp}
\enddata
\tablecomments{$\Phi \times \Delta p/p$ is the product of the phase function $\Phi$ at the time of observation and the relative, rotational linear polarization variability of the object ($\Delta p/p$). The latter is composed of the peak-to-peak rotational linear polarization variability $\Delta p$ normalized by mean linear polarization $p$ during the run. Also listed are the confidence of rejection of the null hypothesis in sigma units along with the reduced $\chi^2$ of each model fit in Table \ref{fourier}.}
\end{deluxetable}

\begin{figure}
\centering
\includegraphics[width=0.47\textwidth]{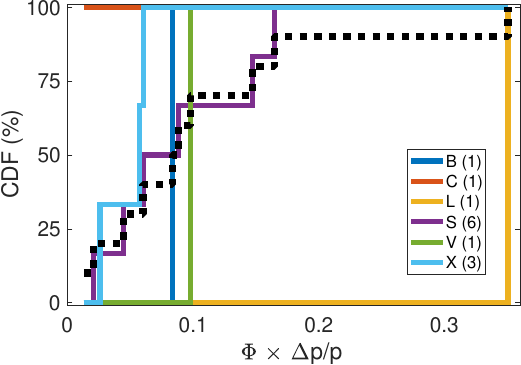}
\caption{Surface heterogeneity versus taxonomic type. Cumulative distribution functions (CDFs) of surface heterogeneity ($\Phi \times \Delta p/p$) for the taxonomic types in this work (Figure \ref{dppsort}). The dashed black curve indicates the CDF for all objects together.}
\label{dpp_tax}
\end{figure}
 
We propose that every increment of 0.05 in $\Phi \times \Delta p/p$ indicates a significant increase in surface heterogeneity (Figure \ref{dppsort}). We measure (1) Ceres, (9) Metis, and (16) Psyche to harbor low surface heterogeneity; (15) Eunomia and (216) Kleopatra to exhibit low-to-moderate heterogeneity; (2) Pallas, (3) Juno, (7) Iris, and (132) Aethra to have moderate heterogeneity; (4) Vesta to harbor moderate-to-high heterogeneity; (6) Hebe to exhibit high-to-extreme heterogeneity; and finally (12) Victoria and (65803) Didymos to have extreme heterogeneity. Indeed, all bodies above save (12) Victoria lie on a fairly continuous spectrum of heterogeneity, where $\Phi \times \Delta p/p$ increases by a factor of $1.248 \pm 0.067$ from one body to the next in Figure \ref{dppsort}. However, the value of $\Phi \times \Delta p/p$ for (12) Victoria is larger by a factor of $2.13 \pm 0.34$ than for (65803) Didymos, the body with the next-largest peak-to-peak heterogeneity. We therefore expect something truly unique to be present on the surface of (12) Victoria and recommend the community investigate. (12) Victoria's extreme linear polarization variations may be a coupling of the relatively large slope of its wavelength-dependent polarization \citep{Belskaya2009} with the clear POLISH2 bandpass in this investigation, but it is unclear why (2) Pallas and (9) Metis do not also harbor such strong polarization variations given their similar absolute value of wavelength-dependent polarization slope \citep{Belskaya2009}.

Our observations of (65803) Didymos suggest that surface heterogeneity is higher for NEOs than for Main Belt Asteroids. This is because the phase function-corrected intrinsic linear polarization variability $\Phi \times \Delta p/p$ of (65803) Didymos, the only NEO in our sample with data quality sufficient to search for such variations, is nearly twice that of (4) Vesta itself. Indeed, linear polarization variability of (65803) Didymos ranks in the 92nd percentile of the objects in our sample, and it is only surpassed by the extreme variability of (12) Victoria.

There are also indications that taxonomy may have a role in surface heterogeneity. Figure \ref{dpp_tax} (left panel) shows cumulative distribution functions (CDFs) of $\Phi \times \Delta p / p$ for each taxonomic type observed by POLISH2. While small number statistics are certainly at play, it is also clear that heterogeneity of S-type bodies (purple CDF) and the full sample (dashed black CDF) are drawn from the same population. Of course, S-types comprise nearly half of the full sample.
 
\subsubsection{\normalsize{The Curious Case of (65803) Didymos}}\label{sec:didy}

The impact of \textit{DART} with (65803) Didymos' satellite Dimorphos at 23:14 UT 26 Sep 2022 presents a unique experiment. VLT 8-m FORS2 and NOT 2.5-m ALFOSC show this impact immediately (within 5.6 h or less post-impact) and potentially irrevocably (at least 81 d post-impact) altered the polarization phase curve of the Didymos system \citep{Bagnulo2023, Gray2024}. Indeed, fractional linear polarization decreased post-impact across $BVRI$ bands from 5.24\% to 4.56\% in $B$ band (absolute difference of 0.68\% and relative reduction by 13 parts in 100) and from 4.35\% to 4.01\% in $I$ band (absolute difference of 0.34\% and relative reduction by 8 parts in 100) between $52.3^\circ < \alpha < 53.6^\circ$ \citep{Bagnulo2023}. FORS2 imaging polarimetry suggests the ejecta cloud was optically thin down to the airless body surfaces after 9.9 h in $B$ band and 7.3 d in $R$ band \citep{Gray2024}. Imaging also shows that fractional linear polarization of the ejecta cloud and tail is perpetually lower than that of the surfaces, which is consistent with the observation that the ejecta cloud reduced system polarization immediately post-impact. However, even after the ejecta cloud dispersed $\sim 23$ d post-impact, Didymos system polarization remained reduced and followed the same polarization phase curve as it had immediately after impact. Thus, it appears that the reduction in Didymos' surface polarization is due to ejecta blanketing the surface of Didymos. Modeling suggests that multiple scattering depolarization is present and caused by impact-generated particles of increased reflectivity. These may be caused by generation of smaller ejecta particles from the regolith or by impact excavation of particles from the subsurface less affected by space weathering \citep{Penttila2024}.

Lick 3-m POLISH2 383 to 720 nm observations of Didymos were obtained from UT 20 to 25 Oct 2022, and they occurred 24 to 29 days post-impact and at the peak phase angle of the apparition ($\alpha \sim 76^\circ$). Fractional linear polarization ranges from $p = 7.28\% \pm 0.20\%$ to $p = 10.59\% \pm 0.26\%$ and is phase locked to Didymos' $\sim 2.26$ h rotation period (Figure \ref{sample1}). This range of fractional polarization is consistent with contemporaneous VLT and NOT $B$, $V$, and $R$ band polarimetry \citep{Gray2024}. Indeed, significant measurement scatter is present in high $\alpha$ Didymos observations by \citet{Gray2024}, who suggest this to be marginal evidence of rotational linear polarization variations in Didymos. We therefore confirm that hypothesis in this work. We expect POLISH2 polarimetry to be dominated by surface polarization of Didymos for the following reasons.

Didymos has $\sim 7$ times the cross-sectional area of Dimorphos. If our measurements of $\Phi \times \Delta p / p = 0.165 \pm 0.025$ (Table \ref{dpp}) are purely due to intrinsic polarization variations across Dimorphos, the satellite must have $\Phi \times \Delta p / p = 1.36 \pm 0.21$. It is not only unphysical for $\Phi \times \Delta p / p > 1$, but it is also unreasonable that Dimorphos would have maximally large polarization variations but the primary body Didymos would have none. Thus, we expect airless body polarization from the Didymos system to be dominated by Didymos itself.

The surface of Didymos was visible 7.3 d or less post-impact, depending on wavelength \citep{Gray2024}. The ejecta cloud was observed to clear $\sim 23$ d post-impact, just before our observations 24 to 29 days post-impact. The ejecta cloud itself was measured to be less polarized than Didymos, as impact caused a dramatic reduction in Didymos-Dimorphos system polarization \citep{Bagnulo2023}. Therefore, given that Didymos dominates airless body polarization over Dimorphos, Didymos' surface was visible during our run, the ejecta cloud was at best only a weak contributor to system polarization during the run, the polarization of Didymos was altered by deposition of material from the ejecta cloud, and that we observe linear polarization modulation tied to Didymos' rotation period with $> 8 \sigma$ confidence (Table \ref{dpp}), we conclude that polarization modulation of the Didymos system during our run is dominated by surface variations on Didymos.

\begin{figure*}
\centering
\includegraphics[width=0.7\textwidth]{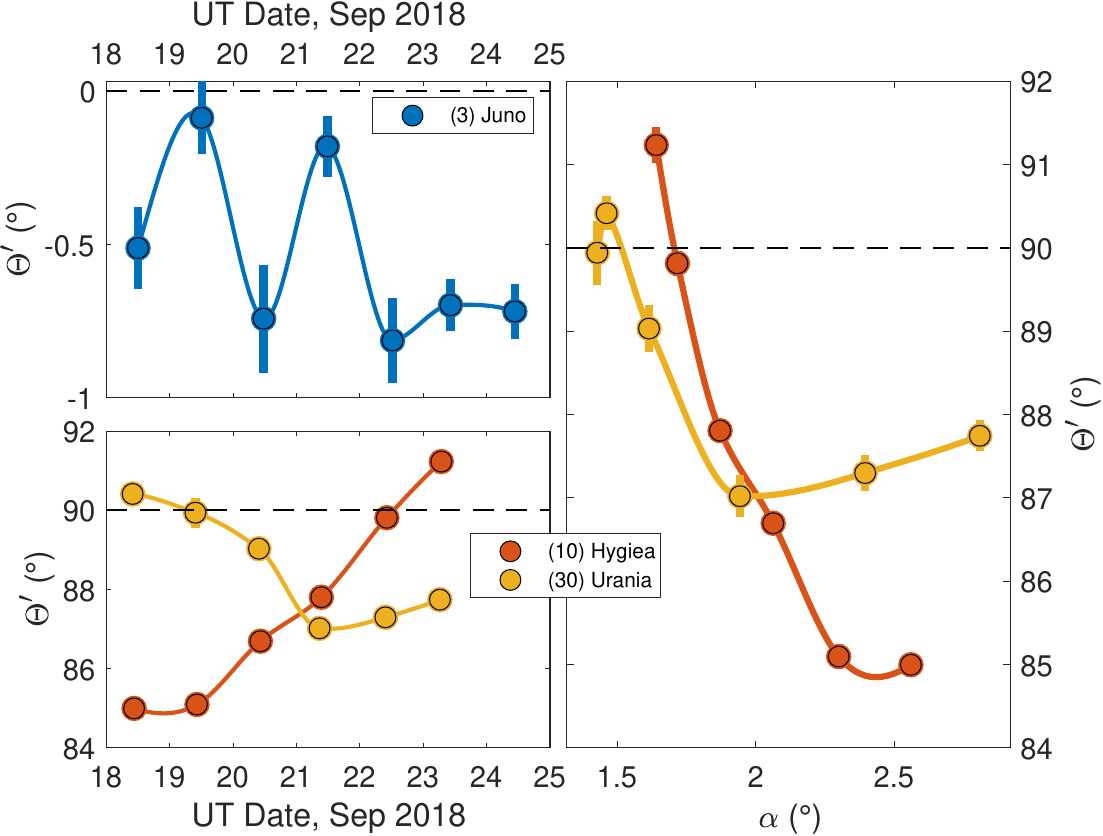}
\caption{Variations in scattering plane-referenced polarization orientation $\Theta'$ observed on three asteroids during the UT 18 to 25 Sep 2018 Lick 3-m POLISH2 run. Note the change in $\Theta'$ scale from object to object. \textit{Top left}: $\Theta'$ as a function of UT date for positive branch observations of (3) Juno obtained at $25\, \fdg 2 < \alpha < 26\, \fdg 4$ that modulate on its 7.2 h rotation period (Figure \ref{sample1}). \textit{Bottom left}: Negative branch observations of (10) Hygiea and (30) Urania obtained for $1\, \fdg 4 < \alpha < 2\, \fdg 8$. Nightly uncertainties in $\Theta'$ are typically $\sigma_{\Theta'} \sim 0\, \fdg1$ and significantly smaller than the size of the data points (Table \ref{alldata}). No change in POLISH2 mounting position occurred during the run, and all three asteroids exhibit either uncorrelated variations in $\Theta' (t)$ (3 Juno versus the others) or anticorrelated variations (10 Hygiea versus 30 Urania) that must be intrinsic to each object. \textit{Right}: $\Theta' (\alpha)$, where the trends of (10) Hygiea versus (30) Urania are correlated for $\alpha < 2^\circ$.}
\label{thetarot}
\end{figure*}

\begin{figure*}
\centering
\includegraphics[width=0.8\textwidth]{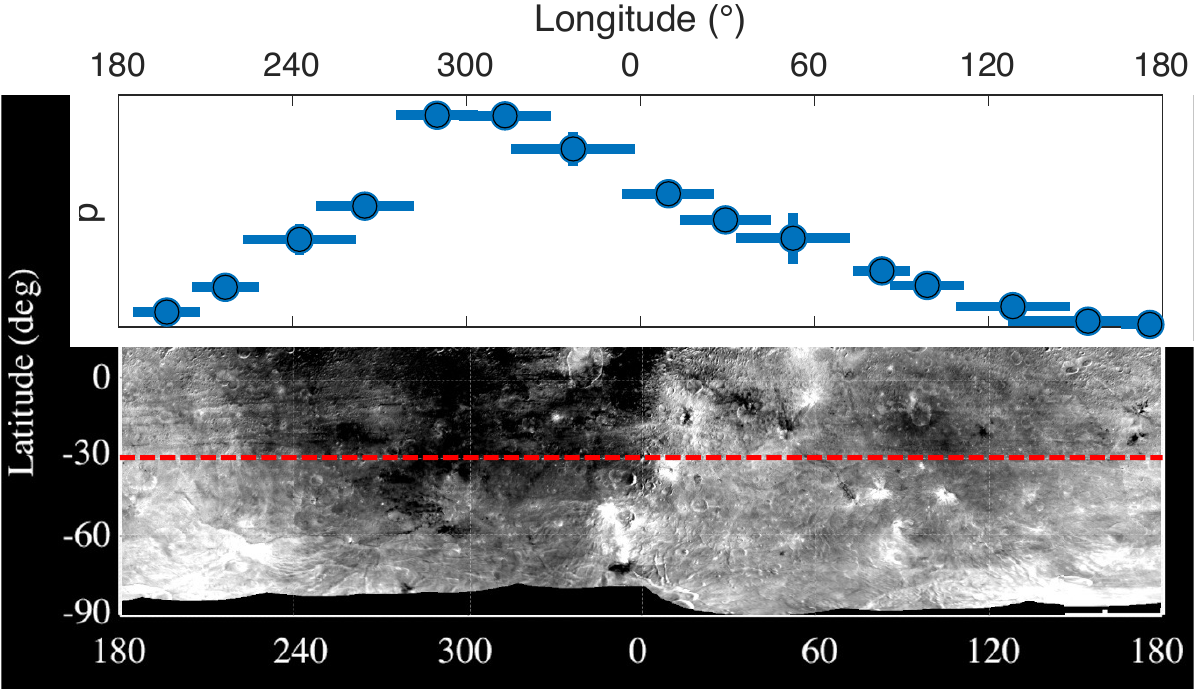}
\caption{Comparison of the \textit{Dawn} albedo map of (4) Vesta \citep{Reddy2012} with POLISH2 linear polarimetry (Figure \ref{sample1}). As expected from the Umov effect, linear polarization increases as the dark Olbers Regio comes into view (longitude $\sim 300^\circ$) and decreases over the bright regions. Sub-observer longitudes for POLISH2 observations are obtained from the JPL Horizons database, and the dashed red line indicates the mean sub-observer latitude of $-30.26^\circ \pm 0.31^\circ$ during POLISH2 observations. The \textit{Dawn} albedo map is modified from \citet{Reddy2012} with permission from AAAS.}
\label{VestaDawnMay18}
\end{figure*}

\begin{figure}
\centering
\includegraphics[width=0.29\textwidth]{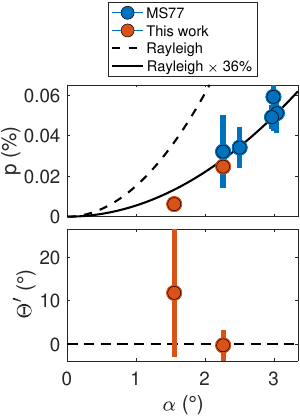}
\caption{POLISH2 383 to 720 nm observations of Uranus obtained at $\alpha = 1.6^\circ$ and $2.3^\circ$ on UT 22 Sep 2018 and 16 Sep 2021, respectively, are compared with measurements from the literature \citep[MS77]{Michalsky1977Uranus}. \textit{Top}: fractional polarization $p$ vs. $\alpha$, which is consistent with Rayleigh scattering with $36.22\% \pm 0.89\%$ efficiency. \citet{Michalsky1977Uranus} observations obtained at $\alpha < 2^\circ$ are generally consistent with zero and are ignored here for clarity. \textit{Bottom}: POLISH2 polarization orientation relative to the plane perpendicular to the scattering plane ($\Theta' = 0.39^\circ \pm 0.72^\circ$), which is consistent with Rayleigh scattering (dashed line). Note \citet{Michalsky1977Uranus} only tabulate $p$ and not $\Theta'$.}
\label{uranus}
\end{figure}

Suppose the pre-impact surface of Didymos were polarimetrically homogeneous ($\Phi \times \Delta p / p \sim 0$) with strong, disk-integrated polarization following the pre-impact phase curve \citep{Bagnulo2023}. Anisotropic settling of depolarizing ejecta from Dimorphos could generate the significant polarization modulation observed by introducing patches of weakly polarized material in the presence of the uniform, strongly polarized surface. However, Figure \ref{sample1} illustrates that patches of low linear polarization occur around rotation phases 0.2 and 0.85. While detailed modeling is required to assess this properly, it seems contrived that patches on nearly opposite sides of the Didymos surface would be preferentially, anisotropically blanketed to cause measured modulation.

We hypothesize a more plausible explanation to be that isotropic blanketing of the entire Didymos surface with ejecta added a uniform, depolarizing layer that softened the larger, pristine variations underneath. Thus, we suggest that not only is the extreme polarization modulation of Didymos in Figure \ref{dppsort} due to intrinsic variations on its surface, but it also underestimates the pre-impact surface heterogeneity. Since impact reduced Didymos' linear polarization by a factor of 13 parts in 100 in $B$ band \citep{Bagnulo2023}, and that POLISH2 measurements are more consistent with VLT 8-m FORS2 $B$ band polarization than $V$ or $R$ bands, we suggest the pre-impact heterogeneity of Didymos to be $\Phi \times \Delta p / p = 0.190 \pm 0.029$ (Figure \ref{dppsort}). This lies within measurement uncertainty obtained on the post-impact body. Since the \textit{DART} impact appears to have slightly dampened the extreme surface heterogeneity of Didymos (Figure \ref{dppsort}), as opposed to causing measured heterogeneity, we hypothesize that NEOs will tend to harbor elevated surface heterogeneity and linear polarization variations with respect to MBAs.
  
\subsubsection{\normalsize{Variations in Polarization Orientation}}

Nearly all scattering plane-referenced polarization orientations $\Theta'$ lie up to a few degrees from $\Theta' = 0^\circ$ or $\Theta' = \pm 90^\circ$ (Tables \ref{alldata} and \ref{alldatarot}, Figure \ref{sample1}), which is unlikely to be an instrumental systematic effect. POLISH2 has mounting pins on the Lick 3-m that enable repeatable clocking on subsequent runs, so the rotational zero point is accurate to $-1\, \fdg08 \pm 0\, \fdg 34$ as measured via strongly polarized standard stars \citep{WiktorowiczNofi2015}. After POLISH2 is mounted on the telescope on the first day of each run, it is kept at the same rotational position for the entire run. In spite of this, significant variations in $\Theta'$ are observed within a single run for some objects (e.g., 3 Juno, 10 Hygiea, and 30 Urania), and some objects observed during the same run do not necessarily share the same offsets in $\Theta'$ (e.g., Oct 2022 observations of 92 Undina and 65803 Didymos). Figure \ref{thetarot} illustrates that the $\Theta'$ trends for (10) Hygiea and (30) Urania, obtained at low phase angle during our 18-24 Sep 2018 run, are anticorrelated with each other in time but are correlated in phase angle for $\alpha < 2^\circ$. The modulation for (3) Juno, obtained during the same run but at high phase angle ($25\, \fdg2 < \alpha < 26\, \fdg 4$), is uncorrelated with either (10) Hygiea or (30) Urania and is tied to asteroid rotation period with $> 8 \sigma$ confidence (Figure \ref{sample1}). It is therefore difficult to identify instrumental systematic effects that generate spurious signatures in $\Theta'$ that are dramatically different from object to object on the same run.

Observed deviations of polarization orientation from the scattering plane and its perpendicular are rare save for a few exceptions such as the Moon \citep{Shkuratov2011} and (21) Lutetia \citep{Belskaya2017}. It is possible that the plane of polarization for each of these objects is rotated slightly away from the scattering plane or its normal due to anisotropic multiple scattering from localized surface features. Spatially resolved polarization orientation variations of the Moon are weak \citep[$\Delta \Theta' \pm 0\, \fdg3$]{Shkuratov2011} and would be difficult to identify in disk-integrated polarization. However, it is possible that these features are more pronounced on asteroids due to their lack of isostasy. That said, the $\Delta \Theta' \sim 6^\circ$ variations in (10) Hygiea require detailed modeling to test this hypothesis.

\subsubsection{\normalsize{Validation from \textit{Dawn} Measurements of (4) Vesta}}

Recall that from the Umov effect, intrinsically dark regolith is strongly polarized, while intrinsically bright regolith is weakly polarized due to depolarization from multiple scattering. Figure \ref{VestaDawnMay18} shows that disk-integrated, Lick 3-m POLISH2 linear polarization of (4) Vesta increases when the dark Olbers Regio is in view and decreases as it rotates out of view, which is consistent with the Umov effect. Thus, disk-integrated polarimetry with POLISH2 does indeed reproduce surface heterogeneity mapped by spacecraft, and the surprisingly large linear polarization variations of (6) Hebe, (12) Victoria, and (65803) Didymos must therefore be caused by strong surface albedo and/or compositional heterogeneity.
 
\subsubsection{\normalsize{Validation from Uranus}}

 Ground-based observations of Uranus and Neptune have uncovered spatially resolved limb polarization via double scattering of sunlight at the limb \citep{Schmid2006}. \citet{Michalsky1977Uranus} detect disk-integrated linear polarization of Uranus with $11 \sigma$ confidence at $\alpha \sim 3^\circ$ using a photoelastic modulator polarimeter similar in principle to POLISH2 (Figure \ref{uranus}). Rayleigh scattering by deep atmospheres such as Uranus must always impart linear polarization oriented perpendicular to the scattering plane ($\Theta' = 0^\circ$) regardless of $\alpha$. In contrast, airless bodies harbor linear polarization oriented parallel to the Sun-body-observer scattering plane ($\Theta' = \pm 90^\circ$) at these low $\alpha$ likely due to coherent backscattering from large regolith particles \citep{Shkuratov1988, Shkuratov1989, Muinonen1989, Muinonen1990, Muinonen2015}.
 
 To validate POLISH2's accuracy, observations of Uranus were obtained at $\alpha = 1.6^\circ$ and $2.3^\circ$ on UT 22 Sep 2018 and 16 Sep 2021, respectively. We detect disk-integrated, fractional linear polarization of Uranus with $p > 0.006\% = 60$ ppm and $7.3\sigma$ confidence, and linear polarization orientation is measured to be $\Theta' = 0.39^\circ \pm 0.72^\circ$ (Figure \ref{uranus}). While \citet{Michalsky1977Uranus} detect fractional linear polarization with higher confidence than in this work, likely owing to the larger $\alpha$ and thereby intrinsic polarization of Uranus at the time of observation, they only provide Stokes $p$ and not $q$, $u$, or $\Theta'$. Our measurements of linear polarization orientation are consistent with $\Theta' = 0^\circ$, which validates POLISH2's ability to accurately measure polarization. Indeed, our discoveries of airless body rotational polarimetric variations (Figures \ref{sample1} through \ref{dppsort}) have larger peak-to-peak values than the uncertainty in our Uranus measurement, $\sigma_p \sim 0.003\% = 30$ ppm, which suggests that POLISH2 discoveries on airless bodies are indeed intrinsic to each body and are not due to instrumental systematic effects.
 
 We combine POLISH2 measurements and those of \citet{Michalsky1977Uranus} obtained at $\alpha > 2^\circ$, as their low phase angle observations are generally consistent with zero. The combination of linear polarization shows that Uranus' fractional polarization $p$ increases with $\alpha$ in a manner consistent with Rayleigh scattering with $36.22\% \pm 0.89\%$ efficiency. That is, we predict peak disk-integrated linear polarization of Uranus at $\alpha = 90^\circ$ to be $36.22\% \pm 0.89\%$ in POLISH2's clear, 383 to 720 nm bandpass. This value lies between Pioneer 11 observations of Jupiter \citep{Smith1984}, Saturn \citep{Tomasko1984}, Titan \citep{Tomasko1982}, and POLISH2 observations of the hot Jupiter HD 189733b \citep{Wiktorowicz2025}. Such strong polarization suggests Uranus' upper atmosphere to be dominated by small particles.
 
\subsection{\normalsize{Discoveries of Optical Circular Polarization}}
 
\subsubsection{\normalsize{Circular Polarization Observations}} \label{sec_circ}

Traditional linear and circular polarization measurement utilizes a hardware swap between half waveplates (for linear polarization measurement) and quarter waveplates (for circular polarization measurement), whose hardware change necessarily changes instrumental crosstalk and ambiguates calibration. This has limited the acceptance of circular polarimetry by the community. On the other hand, not only is POLISH2's linear polarization accuracy transformative for airless body investigations, but the physics of its photoelastic modulators also enables POLISH2 to simultaneously measure linear and circular polarization without any hardware changes \citep{Wiktorowicz2008}. This allows crosstalk, the unwanted, instrumental conversion between linear and circular polarization (and vice versa), to be calibrated to levels below the noise floor \citep{Wiktorowicz2023}. Thus, POLISH2's circular polarization accuracy is truly unprecedented and enables new understanding.

Airless bodies with metalliferous surfaces, such as the Tholen M-type (16) Psyche and (216) Kleopatra, are currently identified by high S- and/or X-band radar albedo \citep{Mitchell1995, Mitchell1996, Neese2012, Shepard2012, Shepard2015, Shepard2016, Shepard2017}. Since not all M-types harbor high radar albedo, not all M-types are thought to be metalliferous. Indeed, the most metalliferous bodies, based on high radar albedo, tend to be M-types with relatively low visual geometric albedo \citep{Magri1999}, hereafter ``visual albedo.'' Interestingly, neither E- nor P-types, nor even M-types with relatively high albedos, appear to be especially metalliferous from radar albedo measurements.

Unfortunately, the decommissioning of Arecibo makes new radar albedo measurements difficult, which advocates for a new technique to identify metalliferous bodies. Vis-NIR spectroscopy is insufficient, because M-type bodies share similar vis-NIR spectra to Tholen E- and P-type bodies (Tholen EMP-type are subsumed in the Bus-DeMeo X-type, \citealt{DeMeo2009}).  Since vis-NIR spectra do not reliably distinguish between EMP-types, they are therefore classified based on visual albedo: E-types have high visual albedo ($p_V > 0.3$), M-types have moderate visual albedo ($0.1 < p_V < 0.3$), and P-types have low visual albedo ($p_V < 0.1$). Recall that the most metalliferous, high radar albedo bodies tend to be M-types at the low end of the visual albedo range, where $0.1 < p_V < 0.18$ \citep{Magri1999}. Note that visual albedo $p_V$ is not to be confused with fractional linear polarization $p$.

\begin{figure}
\centering
\includegraphics[width=0.47\textwidth]{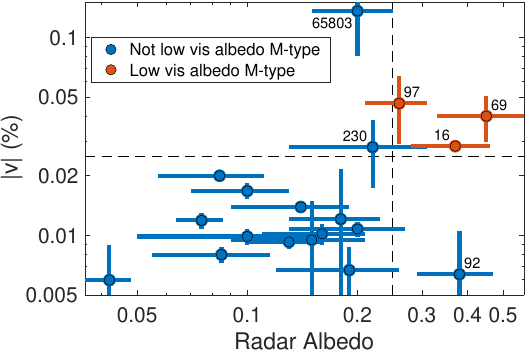}
\caption{Lick 3-m POLISH2 observations of absolute fractional, optical circular polarization $|v|$ as a function of S- or X-band radar albedo. Low visible albedo, M-type asteroids (red points) have both high radar albedo and high optical circular polarization. We detect a clear correlation between circular polarization and radar albedo except for two cases. The circularly unpolarized (92) Undina appears to be a rare M-type asteroid with high visual and radar albedos, and it is possible that (65803) Didymos' circular polarization is contaminated by potentially bright, multiply scattering ejecta from the \textit{DART} impact, or it may be a consequence of its observation at high phase angle. We propose that metalliferous bodies may be identified by absolute fractional circular polarization $|v| > 0.025\% = 250$ ppm and radar albedo $\hat{\sigma}_{OC} > 0.25$ (dashed horizontal and vertical lines, respectively). The relatively large circular polarization of (230) Athamantis must be verified due to its short, 22-min observation.}
\label{radar}
\end{figure}

Metallic powders have been shown in the laboratory to be circularly polarized in the optical, which has been proposed as a potential tool for the identification of metalliferous bodies \citep{Degtyarev1992}. We have discovered that the metalliferous, low albedo M-types (16) Psyche, (69) Hesperia, and (97) Klotho all harbor anomalously high optical circular polarization (Figure \ref{radar}). In contrast, except for Didymos (discussed below), POLISH2 measurements of other bodies that are not low visual albedo M-types tend to have relatively low optical circular polarization, which shows that optical circular polarization may be used to identify metalliferous bodies. We propose that airless bodies may be deemed metalliferous if absolute fractional circular polarization $|v| > 0.025\% = 250$ ppm, which appears to isolate bodies with radar albedo $\hat{\sigma}_{OC} > 0.25$.

K-S tests show that a radar albedo cutoff of $\hat{\sigma}_{OC} > 0.25$ and a circular polarization cutoff of $|v| > 0.025\% = 250$ ppm are equally effective at identifying low visual albedo M-types. Indeed, the probability that metalliferous and non-metalliferous objects have radar albedo drawn from the same population is $0.8\%$, which is the same as the probability that circular polarization of metalliferous and non-metalliferous objects is drawn from the same population. This can be understood intuitively from Figure \ref{radar}, where low visual albedo M-types (red points) and one false positive harbor high radar albedo ($\hat{\sigma}_{OC} > 0.25$). The false positive, the M-type (92) Undina, has high radar albedo but a visual albedo ($p_V = 0.251 \pm 0.014$) that is inconsistent with low visual albedo, M-type asteroids ($0.1 < p_V < 0.18$). Note that (92) Undina's low circular polarization accurately predicts it not to be a low visual albedo, M-type asteroid.

Conversely, Figure \ref{radar} also shows that low visual albedo M-types and one different false positive harbor high circular polarization ($|v| > 0.025\% = 250$ ppm). The false positive, (65803) Didymos, has high circular polarization but is an S-type \citep{deLeon2010, Cheng2018}. Therefore, in addition to having a curious linear polarization signature, Didymos' circular polarization is also puzzling. Given that bright, multiply-scattering ejecta from the \textit{DART} impact reduced Didymos' linear polarization at the time of our Lick 3-m POLISH2 run, it seems plausible that such multiple scattering may also have generated circular polarization significant enough to mimic a metalliferous surface. It is also possible that Didymos' large circular polarization is purely a consequence of high phase angle observations, and that the trend for metalliferous objects in Figure \ref{phase} is not linear but of higher order as measured by \citet{Degtyarev1992} for certain materials. Future observations of additional NEOs at high phase angle will test this hypothesis.

The radar albedo and circular polarization of (230) Athamantis both straddle the proposed $(\hat{\sigma}_{OC}, |v|)$ boundary for metalliferous objects, but it is not a low visual albedo M-type. As this object has the shortest integration time of the objects in Figure \ref{radar}, only 22 minutes on a single night, we propose that follow-up observations are necessary for adjudication. Thus, both radar albedo and circular polarization generally identify low visual albedo, M-type asteroids, as their false negative rate is consistent with zero (all red points in Figure \ref{radar} lie in the upper right quadrant) and their false positive rate is 1 out of 16 (blue points in Figure \ref{radar}).

\begin{deluxetable}{cccccccccc}
\tabletypesize{\tiny}
\tablecaption{Linear Fits to Circular Polarization Phase Curves}
\tablewidth{0pt}
\tablehead{
\colhead{Sample} & \colhead{$|v|$ Slope (ppm$/^\circ$)}	& \colhead{$|v|$ Intercept (ppm)}	& \colhead{Adjusted $R^2$} }
\startdata
Metalliferous     & 8.809(76) 	 & 213.4(1.6)	& 1.000 \\
$\cdots$     	& 18.1(3.9) 	 & 0 (fixed)	& $-$0.338 \\
Non-metalliferous & 1.4(1.2)  	 & 91(21)	& 0.014 \\
$\cdots$     	& 5.7(1.0) 	 	& 0 (fixed)	& $-$0.932
\label{circfit}
\enddata
\end{deluxetable}

\citet{Degtyarev1992} find that the absolute value of circular polarization $|v|$ tends to increase with $\alpha$ for circularly polarized materials in the lab. To test this in POLISH2 data, Figure \ref{phase} (bottom panel) shows airless body circular polarization versus $\alpha$, and the metalliferous bodies (16) Psyche, (69) Hesperia, and (97) Klotho stand out. Circular polarization of these metalliferous bodies depends linearly on $\alpha$, at least at the low phase angles of observation, and the fit coefficients are tabulated in Table \ref{circfit}. In contrast, the slope for the non-metalliferous sample is consistent with zero, which indicates that circular polarization of non-metalliferous asteroids is negligible (adjusted $R^2 \sim 0$ if the $y$-intercept floats and $\sim -0.9$ if it is constrained to zero). Best-fit slopes of the metalliferous and non-metalliferous phase curves differ by $6.0 \sigma$, which shows that the intrinsic circular polarization phase curves of these two samples are distinct. Airless bodies measured in this study tend to be significantly less circularly polarized than the materials measured by \citet{Degtyarev1992}.

We observe circular polarization to change sign from object to object, so results in Figure \ref{phase} and Table \ref{circfit} were obtained using the absolute value of measured circular polarization. This is likely due to a peculiarity of POLISH2's photoelastic modulator system, where Stokes parameter sign is ambiguous for low SNR data \citep{Wiktorowicz2023}. Thus, we cannot test the finding of \citet{Degtyarev1992} that most materials harbor negative, rather than positive, circular polarization.
 
\subsubsection{\normalsize{Circular Polarization Calibration}}

As discussed briefly above, accurate calibration of optical circular polarization tends to be difficult for conventional waveplate polarimeters for many reasons. Optics upstream of the waveplate generate polarization from even an unpolarized beam, and they convert incident linear polarization to circular polarization (``crosstalk''). Waveplate retardance oscillates with wavelength, which causes additional crosstalk across the band. Since conventional polarimeters tend not to be sensitive to both linear and circular polarization simultaneously, it is difficult to measure instrumental linear and circular polarization together. Typically, a half waveplate is inserted for linear polarization calibration and a quarter waveplate is inserted for circular polarization calibration. However, a change to the polarization modulator necessarily changes the instrumental polarization, which leads to ambiguous calibration accuracy across the Poincar\'{e} sphere (combined linear and circular polarization). POLISH2 has been shown to be highly sensitive and accurate to both linear and circular polarization simultaneously, which enables calibration without changes to the optical system \citep{Wiktorowicz2023}. For example, POLISH2 observations of crossed linear polarizers in the lab show that typical stretched-film linear polarizers allow incident Stokes $u$, oriented $45^\circ$ with respect to the polarizer axis, to leak into the output beam with $0.06\%$ efficiency. Such linear polarizers also circularly polarize incident Stokes $u$ with $0.15\%$ efficiency. Incandescent bulbs, fiber sources, integrating spheres, and even cavity blackbodies introduce intrinsic linear and circular polarization of order $0.01\%$ = 100 ppm due to slight asymmetries in the source.

At the Lick 3-m and 1-m telescopes, only the on-axis Cassegrain primary and secondary mirrors lie upstream of the POLISH2 polarizing optics, which minimizes linear (section \ref{instdesc}) and circular telescope polarization contamination. POLISH2 circular polarization calibration is accomplished both by observations of telescope flatfield lamps through right- and left-handed circular polarizers and by observations of the afternoon sky without polarizers. Right- and left-handed circular polarizers enable $v = \pm100\%$ circular polarization (with sign) to be injected into POLISH2, which establishes circular polarization modulation efficiency (effectively polarization gain). Daytime sky observations enable linear to circular polarization conversion to be calibrated by rotating the telescope Cassegrain rotator while collecting data on the bright, linearly polarized sky. Intrinsic linear polarization $q,u$ of the daytime sky modulates sinusoidally during rotation, and it may approach $p = \sqrt{q^2 + u^2} = 100\%$ depending on the phase angle $\alpha$ of the observed patch of sky. Since circular polarization $v$ is invariant under rotation, any sinusoidal modulation of circular polarization during this experiment must be due to instrumental crosstalk. Indeed, \citet{Wiktorowicz2023} measure that POLISH2 converts incident linear polarization to circular polarization with $\sim 1\%$ efficiency. Thus, $p = 100\%$ linearly polarized sky generates spurious circular polarization at the $v \sim 1\%$ level, while airless bodies with $p = 1\%$ to 10\% will only generate spurious circular polarization at the $v = 0.01\%$ to 0.1\% level. Mueller matrix correction of POLISH2 crosstalk leaves circular polarimetric residuals with standard deviation of $\sigma_v \sim 0.02\%$ for incident linear polarization of $p \sim 50\%$ \citep{Wiktorowicz2023}. Thus, POLISH2 delivers circular polarimetric accuracy of 0.0004\% to 0.004\% = 4 to 40 ppm on airless bodies with linear polarization $p = 1\%$ to 10\%, so instrumental contamination lies at least one order of magnitude lower than the detections of circular polarization in Figure \ref{radar}.
 
\subsection{\normalsize{Asteroid Composition and Testing OC Progenitors}}
 
Hayabusa \textit{in situ} measurements of (25143) Itokawa show that ordinary chondrite (OC) meteorites originate from S-type asteroids \citep{Nakamura2011}, and OCs fall into three distinct groups (H, L, and LL). The literature suggests that these three groups represent at least three asteroid collisional families whose recent breakups spawned material that fell to Earth as H, L, and LL chondrites. Asteroid families tend to fall into three WISE W1 albedo groups, which suggests that small compositional differences exist between families \citep{Masiero2014}. Various investigators have proposed that H chondrites arose from (6) Hebe \citep{Gaffey1998}, the Agnia, Merxia, Koronis family \citep{Vernazza2014}, or the Gefion family \citep{McGraw2018}, and that LL chondrites came from the Flora family \citep{Vernazza2008}. The large number of possible parent bodies is largely due to the similarity in 1 and 2 $\mu$m absorption features between these bodies. Thus, it is difficult to spectrally identify specific parent bodies of ordinary chondrites: another technique is needed.

Using approximations of light scattering by surfaces \citep{Muinonen2002a}, investigators find an empirical, linear relation \citep{GilHutton2017} between an asteroid's inversion angle $\alpha_0$ and the mean index of refraction $n$ of its surface. Such a trend has also been found for laboratory samples \citep{Frattin2019}. Index of refraction is related to composition of OCs, where the H, L, and LL OCs are defined by fayalite mole fraction \citep{Nakamura2011}. A linear relation between index of refraction and fayalite mole fraction has been experimentally determined \citep{Bowen1935}, which shows that H, L, and LL OCs will harbor measurably different indices of refraction. Thus, the parent bodies of H, L, and LL OCs should have measurably different polarimetric inversion angles.

\begin{figure}
\centering
\includegraphics[width=0.47\textwidth]{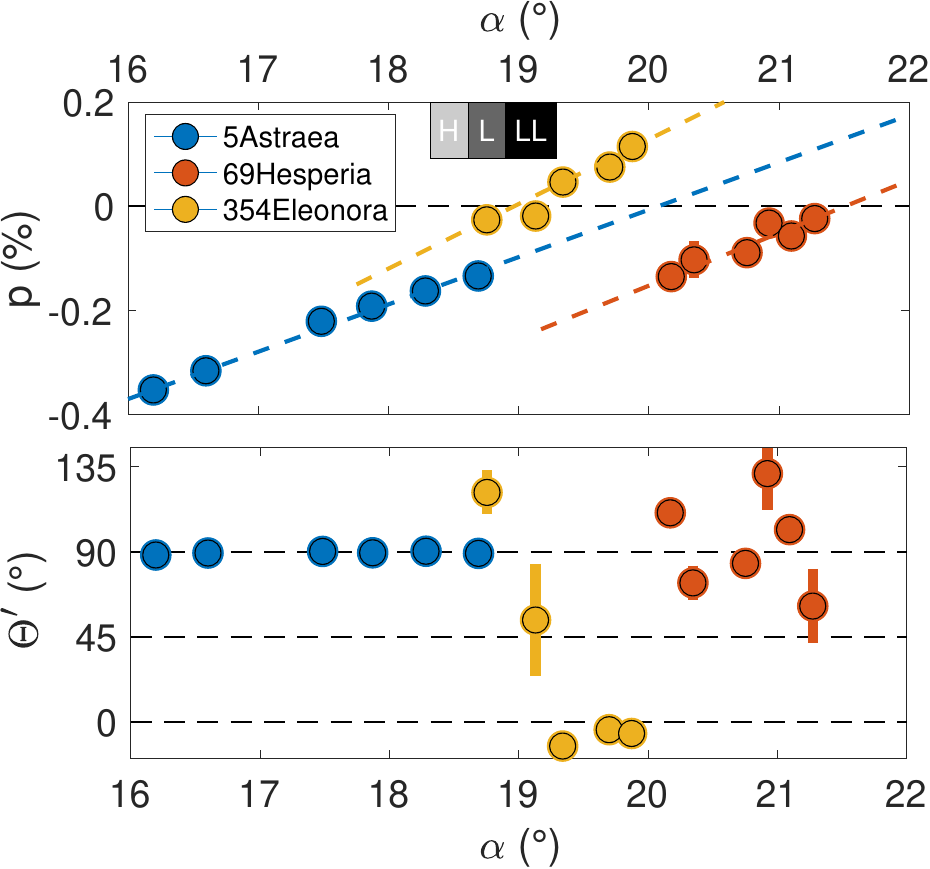}
\caption{Lick 3-m POLISH2 measurements of three asteroids surrounding their inversion angles over a seven-night run. Note the sudden rotation of polarization orientation $\Theta'$ for (354) Eleonora near $\alpha \sim 19^\circ$ (bottom panel, purple points), which causes fractional linear polarization $p$ sign to change (top panel). The expected inversion angles of the parent bodies of H, L, and LL ordinary chondrites are shown at the top of the figure based on laboratory measurements of their refractive indices \citep{Bowen1935, Nakamura2011, GilHutton2017}. Uncertainties are generally smaller than the size of the data points.}
\label{invang}
\end{figure}

Asteroid surface indices of refraction tend to cluster between $1.65 < n < 1.85$ and generate inversion angles of $18^\circ < \alpha_0 < 22^\circ$ (Figure \ref{phase}). \cite{Masiero2009} find that surface index of refraction, and thereby regolith chemical composition, is more important than particle size distribution in determining the location of the inversion angle. Indeed, the L-type asteroid (234) Barbara was discovered to have an anomalously high inversion angle with respect to all other airless bodies \citep{Cellino2006}. This asteroid was later discovered to have 2 $\mu$m absorption indicative of spinel-rich CAIs \citep{Sunshine2008}, whose high index of refraction explains the large inversion angle measured polarimetrically. Thus, the combination of inversion angle polarimetry and NIR spectroscopy provides powerful insight into asteroid surface composition \citep{Vaisanen2020}.

We hypothesize the following: 1) H, L, and LL chondrites each come from recent breakups of S-type collisional families; 2) the differences in chondrite chemical composition arise from compositional differences across collisional families; and 3) family to family differences in polarimetric inversion angles, and therefore of index of refraction and composition, will be consistent with compositional differences between the H, L, LL chondrites. During a single, seven-night Lick 3-m POLISH2 run during relatively poor conditions in February 2020, we measured nightly polarization of three asteroids serendipitously placed near their inversion angles driven by the nightly $\Delta \alpha \sim 0.3^\circ$ change in phase angle (Figure \ref{invang}). This run was primarily scheduled for another scientific investigation, but we measure inversion angles using linear fits to measured polarization for (5) Astraea ($\alpha_0 = 20.07^\circ \pm 0.15^\circ$, S-type), (69) Hesperia ($\alpha_0 = 21.51^\circ \pm 0.18^\circ$, Xk-type), and (354) Eleonora ($\alpha_0 = 18.96^\circ \pm 0.15^\circ$, A-type). While inversion angles are typically fit by polarization phase curves obtained at a large range of phase angles according to the empirical, ``Exponential-Linear model'' $p(\alpha) = A (e^{-\alpha/B} - 1) + C \alpha$ \citep{Muinonen2009}, the reduction in free parameters of a linear fit enable lower uncertainty and higher accuracy in estimated inversion angle.

Clear differences in inversion angle $\alpha_0$ are detected between these three asteroids, which differ with $5.3 \sigma$ confidence between (5) Astraea and (354) Eleonora to $10.9 \sigma$ confidence between (69) Hesperia and (354) Eleonora. Polarization orientation $\Theta'$ is observed to rotate significantly from one night to the next for (354) Eleonora (Figure \ref{invang}, bottom panel). This rotation of the plane of linear polarization is the manifestation of passage through the inversion angle and conclusively demonstrates POLISH2's ability to measure non-zero polarization on the nights before and after the inversion angle geometry. It is apparent that the inversion angles of the S-type asteroid (5) Astraea and X-type (69) Hesperia lie far outside the range expected for OC parent bodies (Figure \ref{invang} top panel), so their mean surface refractive indices are significantly larger than that of OC meteorite samples. Table \ref{ocrej} shows that (5) Astraea and (69) Hesperia may be excluded as OC parent bodies with 4.0 to $13 \sigma$ confidence simply from inversion angle polarimetry alone, depending on the body and type of OC. However, (354) Eleonora's inversion angle implies a surface composition that is consistent with L and LL OCs. Both the index of refraction and inversion angle of a given object will change with wavelength. Since the POLISH2 bandpass in this work is broad, the effective inversion angles of H, L, and LL OCs in Figure \ref{invang} will shift from asteroid to asteroid based on its color.

\begin{deluxetable}{cccccccccc}
\tabletypesize{\tiny}
\tablecaption{Rejection Confidence of Asteroids as OC Parent Bodies}
\tablewidth{0pt}
\tablehead{
\colhead{Asteroid} & \colhead{H Rej Conf ($\sigma$)}	& \colhead{L Rej Conf ($\sigma$)}	& \colhead{LL Rej Conf ($\sigma$)} }
\startdata
(5) Astraea	&  7.8	&  6.4	& 4.0 \\
(69) Hesperia	& 13.1	& 11.9	& 9.0 \\
(354) Eleonora	&  2.4	&  1.0	& 0.5 \\
\label{ocrej}
\enddata
\end{deluxetable}

While these asteroids are not contenders for OC progenitors due to spectroscopic and orbital injection considerations, this pilot study illustrates the capability of high accuracy polarimetry in rapidly vetting future candidates. At visual magnitudes of 10 to 11.5, each asteroid only required 10 minutes of nightly integration time for this study. Since POLISH2 measurement sensitivity scales with photon statistics \citep{Wiktorowicz2008, Wiktorowicz2009, WiktorowiczLaughlin2014, WiktorowiczNofi2015, Wiktorowicz2015_189, Wiktorowicz2023}, full nights of observation enable similar measurement accuracy and discrimination ability for asteroids up to visual magnitude $\sim 15$. Given our finding that measured rotational polarization $\Delta p$ scales linearly with mean linear polarization $p$ (section \ref{linpolvar}), we expect rotational variations not to contaminate the linear polarization phase curve as $p$ vanishes near $\alpha_0$.
 
\subsection{\normalsize{Polarization Opposition Effect}}
 
\begin{figure}
\centering
\includegraphics[width=0.47\textwidth]{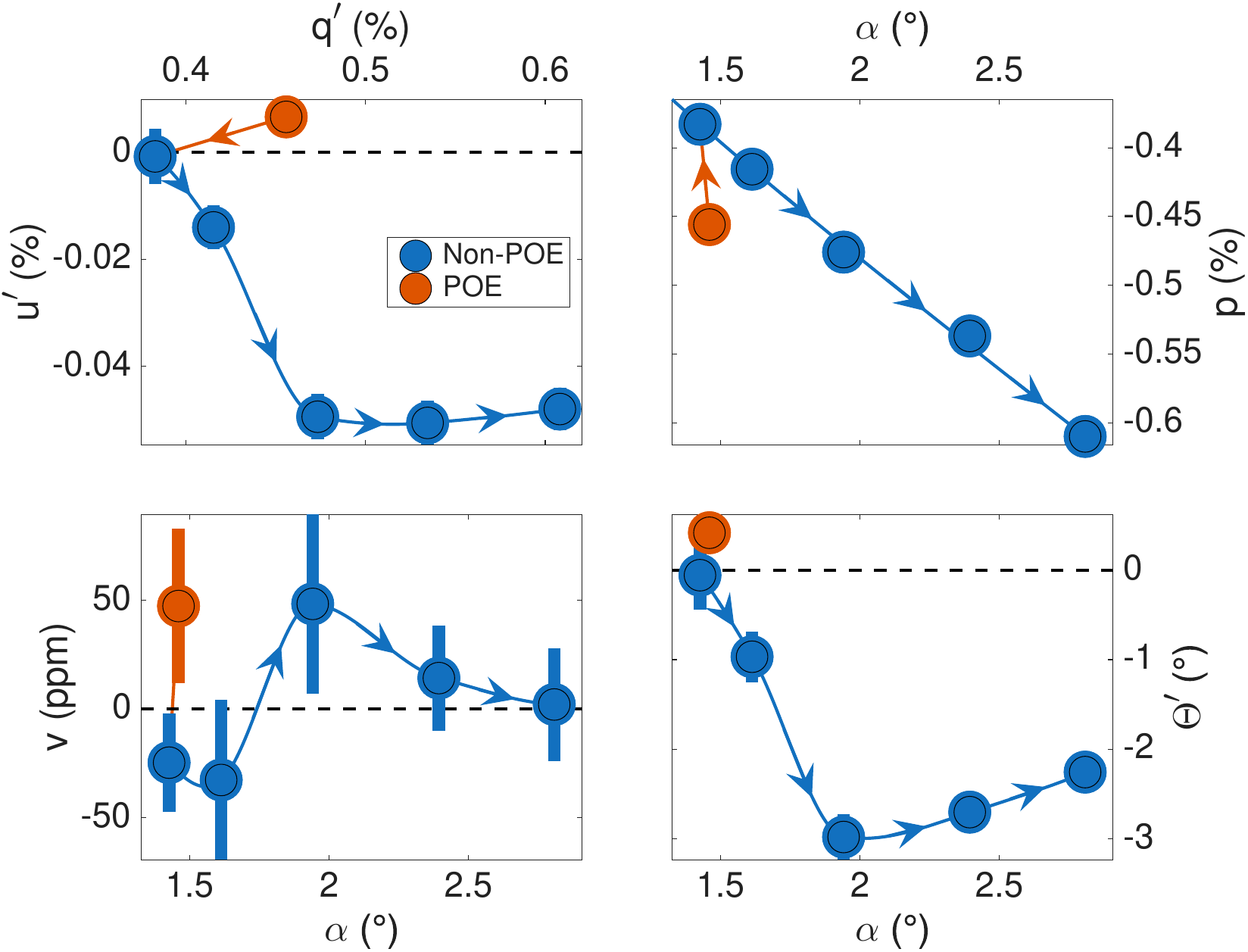}
\caption{Lick 3-m POLISH2 observations during UT 18 to 23 Sep 2018 that may indicate the polarization opposition effect (POE) in (30) Urania. The first night of the run is during the possible POE (red point, UT 18 Sep 2018), and data from successive nights are joined in each panel by curves and arrowheads to show temporal evolution. Some arrowheads are missing for clarity. \textit{Top left}: Nightly linear polarization Stokes $u'$ versus $q'$, rotated to the frame perpendicular to the scattering plane, where data obtained during the night of the possible POE (red point) are inconsistent with the interpolated $(q',u')$ trend (blue curve via Matlab \texttt{makima}) with $11 \sigma$ confidence. Uncertainties are generally smaller than the size of the data points. \textit{Top right}: Fractional linear polarization $p$ as a function of $\alpha$, where data outside the possible POE location (blue points and curve) are fit to the Exponential-Linear model $p(\alpha) = A (e^{-\alpha/B} - 1) + C \alpha$ \citep{Muinonen2009}. \textit{Bottom left}: Fractional circular polarization $v$ as a function of $\alpha$. No significant circular polarimetric manifestation of the possible POE is observed. \textit{Bottom right}: Polarization orientation $\Theta'$ as a function of $\alpha$. A slight excursion of $\Theta' > 0^\circ$ is observed during the night of the possible POE, followed by significant $\Theta' < 0^\circ$ for $\alpha > 1.5^\circ$, though this positive excursion is only inconsistent with $\Theta' = 0^\circ$ with $2.0 \sigma$ confidence.}
\label{poe}
\end{figure}
 
 Recall that coherent backscattering causes polarization orientation $\Theta'$ to rotate by $90^\circ$ near the $\alpha_0 \sim 20^\circ$ inversion angle for airless bodies \citep{Shkuratov1988, Shkuratov1989, Muinonen1989, Muinonen1990, Muinonen2015}. In addition, for $\alpha < 2^\circ$ or thereabouts, nearly perfect backscattering amplifies this interference phenomenon for some bodies. This polarization opposition effect (POE) has been detected on the Galilean satellites and a few asteroids \citep{RosenbushGalilean2005, Kiselev2022}. As the POE is a natural consequence of coherent backscattering, it is perhaps not surprising that measurement of the POE may constrain the following regolith properties: effective size variance, fill factor, and index of refraction \citep{Mishchenko1993}. We present a single outlier in the Stokes $q'$, $u'$ signature for the S-type (30) Urania near $\alpha \sim 1.5^\circ$ that may indicate the presence of the POE for this object (Figure \ref{poe}). Additional measurements are required to test this hypothesis.
  
 \section{Conclusion}
 
 We present a large quantity of linear and circular polarization observations of airless solar system bodies obtained in a clear, 383 to 720 nm bandpass using the POLISH2 polarimeter at the Lick Observatory Shane 3-m telescope \citep{WiktorowiczNofi2015, Wiktorowicz2023} from 2014 to 2022. Observations of (1) Ceres were obtained at the Lick 1-m telescope. We discover rotation phase-locked linear polarization variations with peak-to-peak values $\Delta p > 0.01\%$ in essentially all airless bodies observed, which dramatically increases the number of airless bodies known to show this effect with high accuracy. High phase angle observations of (65803) Didymos show that the peak-to-peak value of polarization modulation $\Delta p$ is correlated with instantaneous fractional linear polarization of the body $p = P/I$. Therefore, relative linear polarization modulation $\Phi \times \Delta p / p$, corrected for the shadowed disk via the phase function $\Phi$ and thereby independent of viewing geometry, is an inherent property of the airless surface and may be compared across airless bodies. Indeed, we discover that (4) Vesta, long displaying the largest surface polarization variations \citep{Degewij1979, Broglia1989, WiktorowiczNofi2015}, is surpassed by the stronger variations in (6) Hebe, (12) Victoria, and (65803) Didymos. Surprisingly, the intrinsic linear polarization variations across (12) Victoria are the largest in the sample and are $\Phi \times \Delta p / p = 3.52 \pm 0.13$ times larger than those of (4) Vesta. While surface albedo heterogeneity drives the amplitude of linear polarization variations in (4) Vesta due to the Umov effect, the unexpectedly large polarization variations across (12) Victoria suggest that compositional heterogeneity may also generate linear polarization variations.
 
 Only two NEOs were observed in this work, and (65803) Didymos is the only one with sufficient rotational phase coverage to search for linear polarization variability. We confirm a marginal detection in the literature of such variability in Didymos \citep{Gray2024} with $> 8 \sigma$ significance, and we show that it harbors the second largest linear polarization variations in our sample after (12) Victoria. The depolarizing effect of \textit{DART}-generated ejecta \citep{Bagnulo2023, Gray2024} implies its pristine variations were even larger, which supports the hypothesis that (65803) Didymos is an aggregate of varying composition. The presence of such strong polarization variations on the only NEO in the sample with sufficient rotational coverage suggests that large rotational linear polarization variations are common among NEOs.
 
 We validate the linear polarimetric accuracy of POLISH2 by detecting increased polarization on (4) Vesta as the dark Olbers Regio rotates into view, which is consistent with the Umov effect. Disk-integrated linear polarization of Uranus' deep Rayleigh scattering atmosphere is measured down to $p = 62 \pm 33$ ppm at $\alpha = 1.6^\circ$, and it consistently displays linear polarization orientation rotated $90^\circ$ from the direction displayed by airless bodies at low phase angles. This is a natural consequence of both Rayleigh scattering, which must be polarized perpendicularly to the Sun-body-observer scattering plane, and coherent backscattering, which causes airless bodies to be polarized parallel to the scattering plane for phase angles less than about $\alpha < 20^\circ$ \citep{Shkuratov1988, Shkuratov1989, Muinonen1989, Muinonen1990, Muinonen2015}. Even low $\alpha$ observations of Uranus show its atmosphere to harbor linear polarization between that of Jupiter \citep{Smith1984}, Saturn \citep{Tomasko1984}, Titan \citep{Tomasko1982}, and the hot Jupiter exoplanet HD 189733b \citep{Wiktorowicz2025}.
 
 In addition to discovering rotation phase-locked linear polarization variations to be common among airless bodies, we discover high radar albedo, metalliferous bodies to be circularly polarized. That is, low visual albedo, M type asteroids harbor circular polarization that is significantly larger than non-metalliferous asteroids. Therefore, we present optical circular polarimetry as a viable, passive technique for the identification of metalliferous bodies.
 
 Composition and compositional heterogeneity of NEOs are significant gaps because of ambiguities and difficulties with traditional techniques. These gaps have significant implications for impactor threat mitigation. Given the single-shot nature of NEO flybys, it is crucial to deploy remote sensing techniques that can gather the most pertinent data during a single flyby. Photometry can determine rotation rate, but shape changes and surface albedo variations are difficult to disentangle. Compositional variations may be completely invisible to photometry. NIR spectroscopy can taxonomically classify NEOs, but rotation-resolved spectral variations are difficult to measure on relatively faint NEOs. Metalliferous surfaces have weak spectral signatures, which requires additional lines of evidence to identify metalliferous bodies. While radar albedo probes metal content, the decommissioning of Arecibo makes it difficult to obtain additional measurements.
 
 High $\alpha$ polarimetric observation of NEOs is a powerful way to probe their surfaces, because both linear polarization and the amplitude of its rotational variations increase with $\alpha$ due to the physics of scattering. Additionally, shadowing of the disk at high $\alpha$ further enhances the detectability of localized linear polarization variations because the diluting flux from the shadowed disk is removed. Indeed, we show that observations of (65803) Didymos at $\alpha \sim 76^\circ$ have peak-to-peak fractional linear polarization of $\Delta p = 3.33\% \pm 0.51\%$, which is nearly two orders of magnitude larger than the median $\Delta p = 0.04 \%$ in our sample.
 
 We measure that circular polarization of metalliferous bodies increases with $\alpha$, which confirms laboratory measurements \citep{Degtyarev1992}. While not expected to be metalliferous, circular polarization of (65803) Didymos is $|v| = 0.136\% \pm 0.055\%$, which is an order of magnitude larger than the median $v = 0.01\%$ of our sample. This implies that metalliferous NEOs will harbor even more pronounced circular polarization at high $\alpha$ than Didymos. Thus, even though NEOs are generally fainter than Main Belt Asteroids, their ability to be observed at high $\alpha$ provides larger linear and circular polarimetric signatures that enhance detectability.
 
 We demonstrate that polarimetry augments spectroscopy and orbital analysis to test airless bodies as ordinary chondrite (OC) progenitors. This is because surface refractive index is correlated with inversion angle $\alpha_0$ \citep{GilHutton2017, Frattin2019}, so measurement of refractive index on airless bodies may be compared to such measurement of OCs. Finally, we detect an anomalous, $11 \sigma$ deviation in linear polarization observations of (30) Urania at $\alpha = 1.5^\circ$ that may be due to the Polarization Opposition Effect. This elusive phenomenon is an extreme consequence of coherent backscattering, whose rotation of polarization orientation at low phase angle has been known for nearly a century. A further investigation dedicated to the search for the Polarization Opposition Effect is required to conclusively identify its presence on (30) Urania and other objects.

\begin{acknowledgments}

This work was supported by NASA SSO grant NNH16ZDA001N-SSO and based on observations obtained at the University of California Observatories/Lick Observatory. We would like to thank Joseph Masiero, Stefano Bagnulo, and Alberto Cellino for their guidance on target selection and scientific context, Ludmilla Kolokolova for discussions about circular polarization, Karri Muinonen for experimental tests of theoretical predictions, Jay Goguen for introducing us to the Polarization Opposition Effect, and Irina Belskaya for discussions about calibration. This work has benefited from the compilation of asteroid radar albedos from William Robert Johnston\footnote{https://www.johnstonsarchive.net/astro/radarasteroids.html}.

\end{acknowledgments}

\facilities{Shane (POLISH2), Nickel (POLISH2)}

\bibliographystyle{apj}
\bibliography{myrefs}

\end{document}